\documentclass[apj]{emulateapj}  
\usepackage{natbib}
\slugcomment{Accepted for publication by the Astronomical Journal}

\newcommand{\cm}{$~{\rm cm^{\scriptscriptstyle -3}}$}
\newcommand{\NblSNIc}{12} 	
\newcommand{\Ngalspec}{17}      
\newcommand{\Nnucgalspec}{9} 	
\newcommand{\Nownspecletter}{Eight}
\newcommand{\Nsloanspecletter}{five}

\newcommand{\NnonTSN}{6}
\newcommand{\hightestzSNIc}{0.14} 

\newcommand{\snbw}{GRB~980425/SN~1998bw}
\newcommand{\sndh}{GRB~030329/SN~2003dh}
\newcommand{\snlw}{GRB~031203/SN~2003lw}
\newcommand{\snaj}{GRB/XRF~060218/SN~2006aj}
\newcommand{\othree}{[\ion{O}{3}]~$\lambda\lambda4959,5007$}
\newcommand{\othreeweak}{[\ion{O}{3}]~$\lambda4363$}
\newcommand{\otwo}{[\ion{O}{2}]~$\lambda3727$}
\newcommand{\ntwo}{[\ion{N}{2}]~$\lambda6584$}

\newcommand{\stwo}{\ion{S}{2}~$\lambda6717$}
\newcommand{\stwol}{\ion{S}{2}~$\lambda6731$}
\newcommand{\snefhost}{UGC~4107}
\newcommand{\sndqhost}{NGC~3810}
\newcommand{\sneyhost}{NGC~7080}
\newcommand{\snaphost}{M~74}
\newcommand{\snbghost}{MCG-05-10-15}
\newcommand{\snjdhost}{MCG-01-59-21}

\newcommand{\sndahost}{UGC11301}
\newcommand{\snfkhost}{SDSS~J211520.09-002300.3}
\newcommand{\snkrhost}{SDSS~J030829.66-005320.1}
\newcommand{\snkshost}{SDSS~J213756.52-000157.7}
\newcommand{\snnbhost}{UGC7230}

\newcommand{\snqkhost}{SDSS~J222532.38+000914.9}
\newcommand{\snihost}{SDSS~J115913.13-013616.0}
\newcommand{\gradkd}{$-$0.49}
\newcommand{\gradm}{$-$0.49}
\newcommand{\gradpp}{$-$0.45}
\newcommand{\centkd}{9.10}
\newcommand{\centm}{9.04}
\newcommand{\centpp}{8.82}
\newcommand{\poskd}{8.62}
\newcommand{\posm}{8.56}
\newcommand{\pospp}{8.38}

\newcommand{\sm}{M$_{\odot}$}
\newcommand{\smy}{M$_{\odot}{\rm yr^{\scriptscriptstyle -1}}$}
\newcommand{\sz}{Z$_{\odot}$}
\newcommand{\kms}{\ensuremath{{\rm km~s}^{-1}}}

\newcommand{\ergscm}{$~{\rm erg~s^{\scriptscriptstyle -1}~cm^{\scriptscriptstyle -2}}$}

\newcommand{\kmsMpc}{${\rm km~s^{\scriptscriptstyle -1}~Mpc^{\scriptscriptstyle -1}}$}
\newcommand{\ergs}{$~{\rm erg~s^{\scriptscriptstyle -1}}$}

\begin{document}

\title{Measured Metallicities at the Sites of Nearby Broad-Lined Type Ic Supernovae and Implications for the SN-GRB Connection\footnote{This paper includes
 data gathered with the MMT Observatory, a joint facility of the
 Smithsonian Institution and the University of Arizona, and with the
 6.5 meter Magellan Telescopes located at Las Campanas Observatory,
 Chile. }}

\author{M.~Modjaz\altaffilmark{1,2}, 
L.~Kewley\altaffilmark{3,4},
R.~P.~Kirshner\altaffilmark{2},
K.~Z.~Stanek\altaffilmark{5}, 
P.~Challis\altaffilmark{2},
P.~M.~Garnavich\altaffilmark{6},
J.~E.~Greene\altaffilmark{3,7},
P.~L.~Kelly\altaffilmark{8},
J.~L.~Prieto\altaffilmark{5}
}

\altaffiltext{1}{Department of Astronomy, UC Berkeley, 601 Campbell Hall, Berkeley, CA, 94704; mmodjaz@astro.berkeley.edu}
\altaffiltext{2}{Harvard-Smithsonian Center for Astrophysics,
 60 Garden Street, Cambridge, MA, 02138; kirshner,pchallis@cfa.harvard.edu.}
\altaffiltext{3}{Hubble Fellow.}
\altaffiltext{4}{University of Hawaii, 2680 Woodlawn Drive, Honolulu, HI 96822;kewley@IfA.Hawaii.Edu.}
\altaffiltext{5}{Department of Astronomy, The Ohio State University, Columbus, OH 43210;kstanek,prieto@astronomy.ohio-state.edu.}
\altaffiltext{6}{Department of Physics, 225 Nieuwland Science Hall, University of Notre Dame, Notre Dame, IN 46556; pgarnavi@nd.edu.}
\altaffiltext{7}{Department of Astrophysical Sciences, Princeton University, Peyton Hall, Ivy Lane, Princeton, NJ 08544-1001; jgreene@astro.princeton.edu.}
\altaffiltext{8}{Kavli Institute for Particle Astrophysics and Cosmology, 2575 Sand Hill Road, MS 29, Menlo Park, CA 94025}

\begin{abstract}

  We compare the chemical abundances at the sites of \NblSNIc\ nearby
  ($z < $\hightestzSNIc) Type Ic supernovae (SN Ic) that showed broad
  lines, but had no observed Gamma-Ray Burst (GRB), with the chemical
  abundances in 5 nearby ($z < $0.25) galaxies at the sites of GRB
  where broad-lined SN Ic were seen after the fireball had faded. It
  has previously been noted that GRB hosts are low in luminosity and
  low in their metal abundances. If low metallicity is sufficient to
  force the evolution of massive stars to end their lives as GRB with
  an accompanying broad-lined SN Ic, then we would expect higher metal
  abundances for the broad-lined SN Ic that have no detected GRB. This
  is what we observe, and this trend is independent of the choice of
  metallicity calibration we adopt, and the mode of SN survey that
  found the broad-lined SN. A unique feature of this analysis is that
  we present new spectra of the host galaxies and analyze all the
  measurements of both samples in the same set of ways, using the
  galaxy emission-line measurements corrected for extinction and
  stellar absorption, via independent metallicity diagnostics of
  Kewley \& Dopita (2002), of McGaugh (1991) and of Pettini \& Pagel
  (2004). In our small sample, the boundary between galaxies that have
  GRB accompanying their broad-lined SN Ic and those that have
  broad-lined SN Ic without a GRB lies at an oxygen abundance of
  12+log(O/H)$_{\rm{KD02}}\sim$ 8.5, which corresponds to 0.2$-$0.6
  \sz\ depending on the adopted metallicity scale and solar abundance
  value. Even when we limit the comparison to SN Ic that were found in
  untargeted supernova surveys, the environment of every broad-lined
  SN Ic that had no GRB is more metal rich than the site of any
  broad-lined SN Ic where a GRB was detected.
\end{abstract}

\keywords{galaxies: distances and redshifts --- gamma
rays: bursts supernovae: general galaxies: abundances  \\ }

\section{INTRODUCTION}\label{intro_sec}

We seek clues to the stellar origin of long-duration Gamma Ray Bursts
(GRBs) by comparing the chemical abundances at the sites of the broad-
lined Type Ic supernovae (SN Ic) that accompany some GRB with the
broad-lined SN Ic that have no observed GRB. In nearby GRB, after the
afterglow has faded, if the spectrum of the underlying event is
observed, it is that of a broad-lined SN Ic (e.g.,
\citealt{galama98,stanek03,hjorth03,modjaz06}). These are supernovae
whose spectra show no trace of hydrogen or of helium and whose line
widths approach 30,000 \kms\ (e.g., \citealt{galama98,patat01,pian06}).
Type Ic supernovae with broad lines are also seen without an
accompanying GRB. We have studied these objects, and their hosts, to
try to learn more about the conditions that lead a massive star to
have the special kind of core collapse that leads to the formation of
a jet and a GRB (see \citealt{woosley06_rev} for a review).

Some models of core collapse in massive stars produced to match these
spectra have large kinetic energies ($E_K > 10^{52}$ erg;
\citealt{iwamoto98,mazzali03}), and thus, have led to the use of the
term "Hypernova". However, since the computed $E_K$ are
model-dependent \citep{maeda06} and for some cases not necessarily
larger than those of normal core-collapse SN \citep{mazzali06_06aj}, we
prefer to use names that reflect the directly observed properties,
without any implication on the overall energy budget for the
event. Thus, we call them ``broad-lined SN Ic'', and abbreviated as
``SN Ic (broad)'', and for comparison, we henceforth call and
abbreviate broad-lined SN Ic associated with GRB as ``SN Ic (broad \&
GRB)'' (see \S~\ref{sample_sec}).

A number of studies have been conducted on the environments of GRBs to
help constrain their progenitors. Observations of host galaxies of
nearby GRBs ($z<$0.25) indicate that they are faint, blue,
star-forming galaxies with low metallicities\footnote{Here we refer to
the term ``metallicity'' as represented by the gas-phase oxygen
abundance, and use these two terms interchangibly. See
\S~\ref{metal_subsec} for details.}
\citep{prochaska04,sollerman05,modjaz06,stanek07,thoene06,wiersema07,margutti07}.
Studies of GRB hosts at cosmological distances draw a similar
conclusion (\citealt{fruchter99,lefloch03,fynbo03,fruchter06}): GRB
hosts tend to be consistently dimmer and more irregular than galaxies
that host core-collapse SN at comparable redshifts and GRBs appear
concentrated in the highest surface brightness areas of their blue
hosts. In addition, \citet{kewley07} and \citet{brown06_metal} have
discovered a number of extremely metal poor galaxies that do not
follow the regular galaxy luminosity-metallicity (hereafter $L-Z$)
relationship \citep{garnett02,tremonti04}, just like the nearby GRB
host galaxies. For the same $M_B$, they are more metal-poor than
Irregulars, normal Blue Compact Galaxies and normal star-forming
galaxies (see also Figure 3 in \citealt{savaglio06_conf} for a similar 
trend for hosts of higher $z$ GRBs). Thus, based on observations,
\citet{fruchter06} and \citet{stanek07} argue that low metallicity is
required for GRB progenitors, in line with theoretical GRB models that
suggest rapidly rotating massive stars at low metallicity
\citep{hirschi05,yoon05,woosley06_z,langer06} as likely
progenitors. Low metallicity seems to be a promising route for some
stars to avoid losing angular momentum from mass loss
\citep{vink05,crowther06}. High angular momentum and a massive core
are key ingredients for producing a GRB jet. If the abundances in the
star are low enough, then hydrogen and helium can mix into the burning
zones of the star, leading to a star with low hydrogen and helium
abundances and a large core mass just before explosion
\citep{hirschi05,yoon05,woosley06_z,langer06}. This mechanism for
producing a gamma-ray burst seems plausible for producing a
broad-lined SN Ic at the same time, as observed.

However, the SN Ic that are observed to erupt without a GRB may
provide essential clues to complete this picture.  If low metal
abundance always accompanies jet formation in the core collapse of
massive stars, then broad-line SN Ic intrinsically without GRBs should
be found in metal-rich galaxies.  This is the hypothesis we test here,
using the available list of broad-lined SN Ic and our own uniform
determinations of the local metal abundance. Though the present sample
is small, it is intriguing, and we look forward to future discoveries
that can test the conclusions presented here.

After the discovery of SN~1997ef, around 15 additional broad-lined SN
Ic were discovered, which have spectroscopic properties similar to the
broad-lined SN Ic-GRB when studied in detail. Studying the full class
of broad-lined SN Ic may help us understand the mechanism for
producing jet-driven explosions. Since these SN are more frequent than
nearby GRBs, we may gain clues more rapidly than waiting for the
detection of the next nearby GRB. Three of the nearby SN-GRB have
gamma-ray energies 2$-$4 orders of magnitude lower compared to
cosmological ones and might belong to a different class of GRBs that
is much more frequent (by a factor of $10^2 - 10^3$), but not
observable at high $z$ due to their low gamma-ray energies
\citep{cobb06,liang06,guetta06}. Thus, it is still worthwhile to
investigate the specific conditions which give rise to low-$z$ GRBs,
even if those insights cannot be directly applied to cosmological,
high-energy GRBs.

Previous work has investigated the metallicity dependence of GRBs;
however these studies \citep{wolf07} are based on indirect measures of
galaxy metallicity. Not only do these relations contain significant
scatter \citep{tremonti04}, but there is evidence that the $L-Z$
relation is no longer linear at very low metallicity for certain
galaxies \citep{kewley07,brown06_metal,kniazev03}. In this paper, we
present the $measured$ metallicities of a statistically significant
sample of broad-lined SN Ic environments, derived in the same fashion,
with which the SN-GRB host sample should be and will be compared. This
study constitutes the first of its kind for core-collapse supernovae.

In \S~\ref{GRBSNsample_sec} we summarize the sample of broad-lined SN
Ic with GRBs from the literature, while in \S~\ref{sample_sec} we
introduce our sample of local broad-lined SN Ic and their hosts and discuss the role of selection effects in \S~\ref{sampleselection_sec}. We
then present the spectroscopic observations in
\S~\ref{obs_sec} and derive galaxy parameters in
\S~\ref{hosts_sec}. We discuss our results and their implications in
\S~\ref{results_sec} and summarize in \S~\ref{summary_sec}.  In this
paper, we adopt the following values of the cosmological parameters:
$H_o = 70$~\kmsMpc, $\Omega_m = 0.3$, and $\Omega_{\Lambda} = 0.7$.

\section{SN Host Galaxy Samples}
\subsection{Sample of Broad-lined SN Ic with GRBs and their Hosts}\label{GRBSNsample_sec}

There are currently four nearby secure cases of direct SN-GRB
associations: the temporal and spatial coincidence between SN~1998bw
and GRB~980425 \citep{galama98} and the metamorphosis of the GRB
afterglow spectra into that of a supernovae for the following SN-GRB:
\sndh\ \citep{stanek03,hjorth03,matheson03}; \snlw\
\citep{malesani04,mazzali06_03lw}; \snaj\ \citep{modjaz06,pian06,mazzali06_06aj,campana06,mirabal06,sollerman06,cobb06}.
We also include the host of XRF~020903, that had a clear supernova
signature in its light curve \citep{bersier06} and in its afterglow
spectrum \citep{soderberg05}, but we do note that this SN confirmation
has a lower degree of certainty than the other GRB-SN
associations.\footnote{There is some evidence of spectroscopically
confirmed supernovae for other GRBs, however with less confidence
(\citealt{dellavalle03}; see \citealt{woosley06_rev} for a full
list).}  The nearby SN connected with GRBs have been well-observed and
in all cases, their host-galaxy properties and emission line fluxes
measured. We list in Table~\ref{GRBSNsample_table} the SN-GRB sample,
and the references include published host galaxy emission line fluxes
which we directly use when deriving metallicities in
\S~\ref{metal_subsec}. While there are certain conditions that might
favor detecting certain SN spectroscopically in GRB afterglows (e.g., good
burst localization, weak GRB afterglow, low redshift), a detailed
analysis is beyond the scope of this paper, and we refer the reader to
the discussion in \citet{woosley06_rev}.

\subsection{Sample of Broad-lined SN Ic and their Hosts}\label{sample_sec}

In our study, we include the host galaxies of the four broad-lined SN
Ic which are well-documented in the literature and whose properties
have been observed and modeled: SN~1997ef and SN~1997dq
\citep{iwamoto00}, SN~2002ap (e.g., \citealt{gal-yam02_02ap,mazzali02_02ap,foley03}), and SN~2003jd
\citep{mazzali05_03jd}. We also include SN~2003bg, which was
classified as a broad-lined SN Ic \citep{filippenko03_03bg}. It then
developed P Cygni absorption features of hydrogen
\citep{hamuy03_03bg}, but did not display any H emission in late-time
spectra as expected in Type II SN (M. Hamuy et al., in prep.). Thus,
we count SN~2003bg as a broad-lined SN Ic, with some hydrogen possibly
due to ISM interactions \citep{soderberg06_03bg}. In addition, we
searched the International Astronomical Union Circulars (IAUCs)\footnote{http://cfa-www.harvard.edu/iau/cbat.html} for announcements
of broad-lined SN Ic. We list the sample for which we obtained data in
Table~\ref{sample_table}. Column (1) indicates the supernova, column
(2) its host galaxy, columns (3) and (4) the Hubble type (in the RC3
system) and redshift of the host galaxy as listed in the NASA/IPAC
Extragalactic Database (NED) or as determined by the SDSS pipeline for
SDSS galaxies. Columns (5) and (6) list the offsets in RA and DEC of
the SN from the host galaxy nucleus. The discovery IAUC and the
circulars with the spectroscopic SN identification are listed in
column (7). In column (8) we list the fashion in which the broad-lined
SN Ic were discovered: ``T'' stands for cases where the SN host galaxy
was targeted, while ``Non-T'' SN hosts were not.  The latter group consists
of SN that were found in rolling, wide-field and non-targeted searches, or in
background-galaxies of fields with targeted galaxies. The discovery
method is important for assessing the significance of selection
effects in our sample. We discuss the implications and significance of
column (8) in the following section in more detail.

\subsection{Controlling for Selection Effects}\label{sampleselection_sec}

GRBs and SN are usually found in dramatically different ways. GRBs are
discovered in wide-field, nearly all-sky searches with Gamma-Ray
satellites such as BATSE, HETE or Swift, whose afterglow studies have
lead to the discovery of SN in certain GRBs.

Most traditional searches for nearby SN, however, are galaxy-targeted
searches. The successful Lick Observatory SN Search (LOSS,
\citealt{filippenko01}) and many amateur SN searches possess a
relatively small field-of-view (FOV, $8{\farcm}7 \times 8{\farcm}7$
for LOSS) and have a database of galaxies that they monitor nightly.
Thus, they include well-known and inevitably, more luminous galaxies
(e.g., \citealt{li01,gallagher05,mannucci05}). In particular, the LOSS
database of 10341 galaxies with measured $M_B$-values has a mean
(median) value at the galaxy magnitude of $M_B=-19.9$ ($M_B=-20.1$)
mag with a standard deviation of 1.3 mag (J. Leaman, private
communication). We are aware that selection effects can be introduced
by different methods of discovery. This is why we have worked hard to
include SN that were found in host galaxies that had not been targeted
for search. We believe they provide a better match to the host
galaxies that are selected by the appearance of a GRB.

Those \NnonTSN\ broad-lined SN were found either via the Sloan Digital
Sky Survey-II (SDSS-II) SN
survey\footnote{http://www.sdss.org/supernova/aboutsprnova.html}, or
via the Texas SN
Survey\footnote{http://grad40.as.utexas.edu/~quimby/tss/index.html}
\citep{quimby06_thesis} that are both rolling searches with a large
FOV (1$-$3 square degrees) and thus can be considered galaxy-impartial
surveys. In other cases, they were found in background galaxies of
LOSS target fields (W. Li, private communication), and thus, can also
be considered SN with non-targeted host galaxies. Assuming the sample
of SDSS star-forming galaxies from \citet{tremonti04} is
representative of the galaxies surveyed by the SDSS-II SN search, the
mean (median) galaxy luminosity is $M_B\sim-19.0$ ($M_B\sim-19.1$)
mag, i.e., one magnitude fainter than the mean value for the LOSS
sample. Moreover, the untargeted searches employ image-subtraction
techniques and thus are sensitive to SN occuring at the center of
galaxies.

For completeness, we analyze and discuss spectra of all SN Ic (broad).
But when directly comparing to the sample of SN Ic (broad \& GRB), we
only include SN Ic (broad) found in the same, non-targeted fashion in
order to minimize discovery selection effects. The sample size of SN
Ic (broad) that are free of galaxy-selection effects is comparable to
that of SN-GRB.

\subsection{GRB Non-Detections}

Here we investigate whether the broad-lined SN Ic in
Table~\ref{sample_table} had a detected GRB.

\citet{wang98} suggested that SN~1997ef and SN~1997dq were associated
with GRB~971115 and GRB~971013, respectively. However, their spatial
and, more stringently, their temporal concurrences are much weaker than
for \snbw~(for detailed discussion see \citealt{li06_grbsn} and
\citealt{woosley06_rev}). Thus, claims of GRB associations for
SN~1997ef and SN~1997dq are much less compelling than those for cases
such as \snbw. Furthermore, no GRB was found to be associated with
SN~2002ap (\citealt{hurley02}).

\citet{mazzali05_03jd} have argued for an off-axis GRB in SN~2003jd
via late-time spectral signatures of an aspherical explosion seen from
along the equatorial plane.  On the other hand,
\citet{soderberg06_radioobs} present a search for late-time radio
emission of a sample of 68 SN Ib/c and argue against a GRB-connection
for SN~2003jd, as well as for all non-SDSS SN of our sample (except
SN~2005nb). We discuss if SN~2003jd had an off-axis GRB in more detail
in \S~\ref{caveats_subsec}, along with the broader question of viewing
angle effects.

For the rest of the broad-lined SN Ic in our sample, we consulted the
GRB localizations\footnote{See http://www.mpe.mpg.de/$~$jcg/grbgen.html}
and did not find any significant spatial (within $\pm$ 2
$\deg$) nor temporal (within $\pm$ 20 days) concurrences. Thus, we
assume that no GRBs were observed for those broad-lined SN
Ic.


\section{Spectroscopic Observations}\label{obs_sec}

\subsection{CfA Data}

\Nownspecletter~of the total of \Ngalspec~host-galaxy spectra were
obtained with the 6.5 m Clay Telescope of the Magellan Observatory
located at Las Campanas Observatory (LCO), with the 6.5 m Multiple
Mirror Telescope (MMT) and the 1.5 m Tillinghast telescope at the Fred
Lawrence Whipple Observatory (FLWO).  The spectrographs utilized were
the LDSS-3 (Mulchaey \& Gladders 2005) at LCO, the Blue Channel
\citep{schmidt89} at the MMT, and FAST \citep{fabricant98} at the FLWO
1.5m telescope. All optical spectra were reduced and calibrated
employing standard techniques (see e.g., \citealt{modjaz01}) in
IRAF\footnote{IRAF is distributed by the National Optical Astronomy
Observatory, which is operated by the Association of Universities for
Research in Astronomy, Inc., under cooperative agreement with the
National Science Foundation.} and our own IDL routines for flux
calibration.


\begin{figure}[!ht]
\plotone{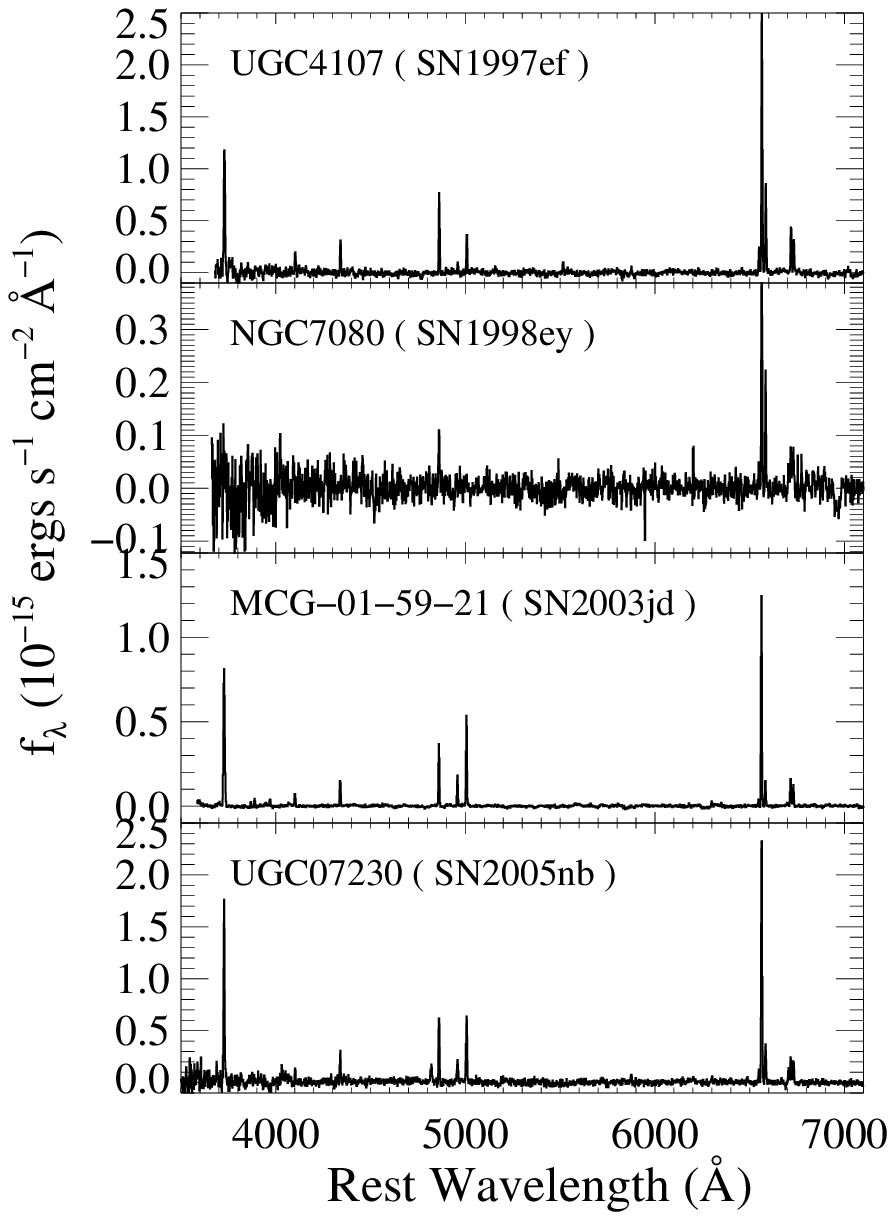}
\caption{\singlespace Host-galaxy spectra of four local broad-lined SN Ic at the SN
position, with the SN spectrum subtracted. The host-galaxy names and
corresponding SN are given in the caption. They were de-redshifted
using values listed in Table~\ref{sample_table}. See \S~\ref{obs_sec}
and Table~\ref{obs_table} for details.}
\label{SN_subgals4_fig}
\end{figure}

Extraction of the spectra were done using the optimal weighting
algorithm \citep{horne86}, and wavelength calibration was accomplished
with HeNeAr lamps taken at the position of the targets. Small-scale
adjustments derived from night-sky lines in the observed frames were
also applied. The spectra were always taken at or near the parallactic
angle \citep{filippenko82} and at low airmass. Comparison of $(B-V)$
colors of SN derived from spectroscopy using the same telescopes and
instruments yield consistency with SN photometry at the ~5\% level
\citep{matheson07}; thus, we are confident that our relative flux
calibration is accurate to $\sim 5\%$ across 4000$-$6500 \AA\, and
most likely across the rest of the covered wavelength range. Telluric
lines were removed following a procedure similar to that of
\citet{wade88} and \citet{matheson00_93j}. The final flux calibrations
were derived from observations of spectrophotometric standard stars.

\begin{figure}[!ht]
\plotone{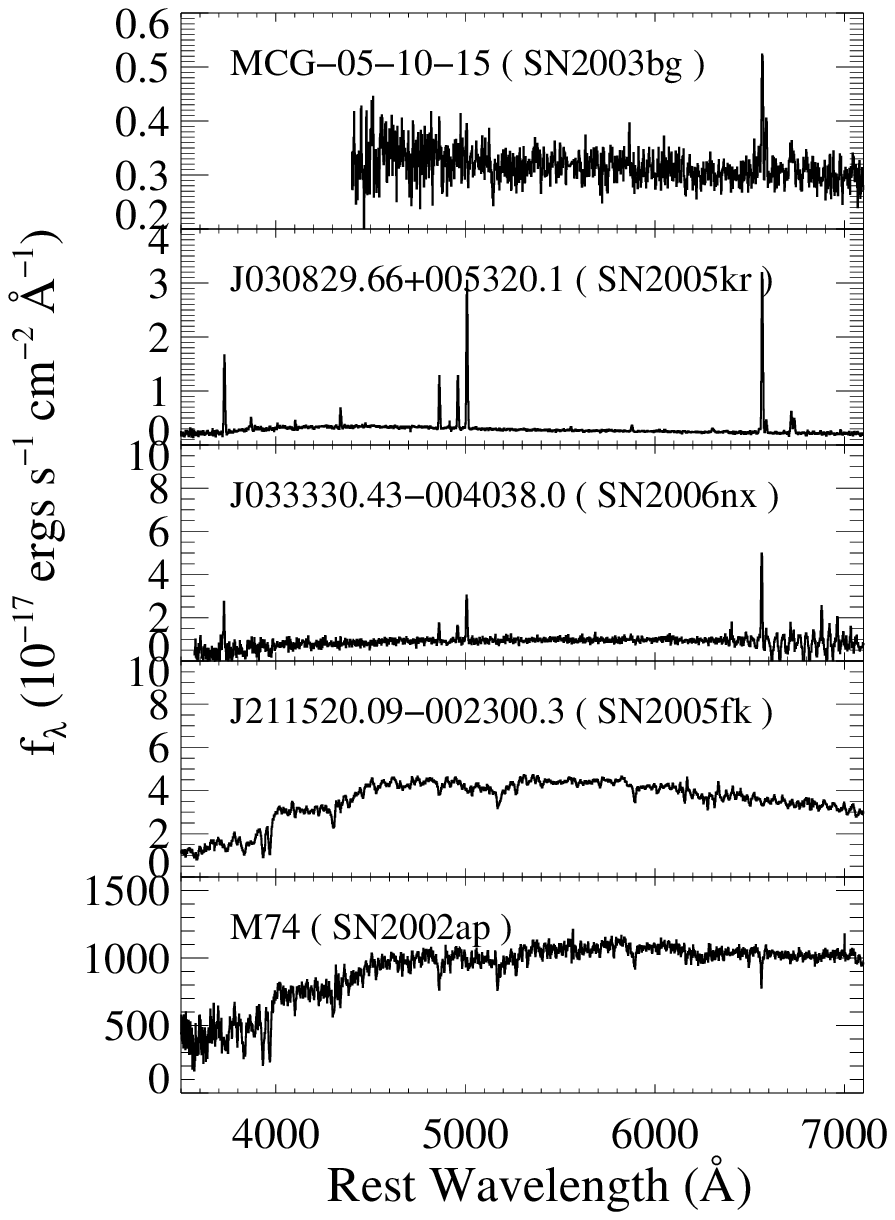}
\caption{\singlespace Central host-galaxy spectra of five local broad-lined SN Ic
observed with Magellan+LDSS3 and the FLWO1.5m+FAST. The host-galaxy names and corresponding
SN are given in the caption. They were de-redshifted using values
listed in Table~\ref{sample_table}. See \S~\ref{obs_sec} and
Table~\ref{obs_table} for details.}
\label{SN_nucgals5_fig}
\end{figure}

In Table~\ref{obs_table} we list the details of our spectroscopic
observations. Exposures over multiple nights were averaged for
SN~1997ef and SN~2003jd to increase the signal-to-noise ratio
(S/N). In four cases (\snefhost, \sneyhost, \snjdhost, \snnbhost, i.e. hosts of
SN 1997ef, 1998ey, 2003jd and SN 2005nb respectively), a spline fit to the
continuum was subtracted, in order to eliminate the supernova
contribution. This procedure is adequate to study the emission line
component of the host galaxy at the position of the SN. Since the SN
features are very broad ($\sim$ 10,000$-$30,000 \kms) compared to the
nebular galaxy emission lines (200$-$400 \kms), there are no
significant errors ($\la$ 5 \%) in the emission line measurements due
to interpolation errors. 
Figure~\ref{SN_subgals4_fig} shows the emission-line spectra of four
host galaxies at the position of the SN, with the SN spectrum
subtracted.  In Figure~\ref{SN_nucgals5_fig}, we present central
spectra obtained with Clay+LDSS3 and FLWO1.5m+FAST.

\subsection{Archival Data}

\begin{figure}[!ht]
\plotone{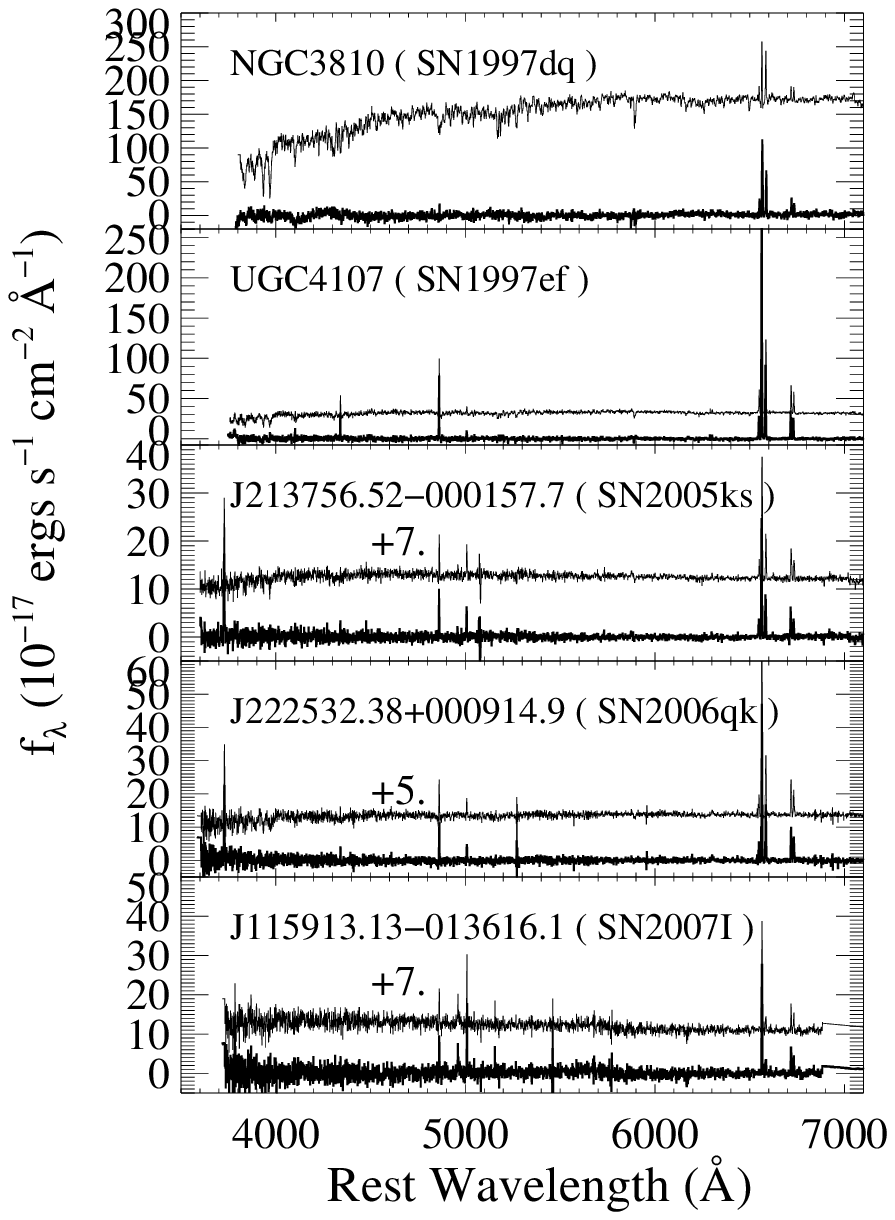} 
\caption{\singlespace  Central host-galaxy spectra of five local broad-lined SN Ic
taken from the SDSS database plotted with (\emph{top}) and without
(\emph{bottom}) stellar contribution (see \S~\ref{pca_subsec} for details). Any offsets by an additive constant done for
clarity purposes are indicated. Note that weak nebular emission lines
become more apparent in stellar-continuum-subtracted spectra. The host
galaxy names and corresponding SN are given in the caption. They were
de-redshifted using values listed in Table~\ref{sample_table}. See text
for details.}

\label{SN_sloangals5_fig}
\end{figure}

\Nsloanspecletter\ central host-galaxy spectra were taken from the
Fifth Data Release of the Sloan Digital Sky Survey (SDSS;
\citealt{york00,stoughton02})\footnote{\singlespace We used the
Princeton spectral reductions (http://spectro.princeton.edu/). We note
that for two of the \Nsloanspecletter\ SDSS galaxy spectra, the
Princeton reductions give fluxes larger by a factor of 2 than the
official data release at
http://cas.sdss.org/dr5/en/tools/explore/. This difference is merely
due to a different flux normalization convention between the SDSS
database (spectrum normalized to the fiber magnitude) and the
Princeton database (spectrum normalized to the PSF magnitude), and
does not imply different reduction steps (J. Frieman, private
communication). Nevertheless, this normalization convention does not
change our metallicity determinations as they are based on line flux
ratios.}. The spectra were acquired with the SDSS 2.5m telescope using
fiber-fed spectographs, with the fiber subtending a diameter of
3${\arcsec}$ (for a full discussion of aperture effects see
\citealt{tremonti04}). Exposure times had been chosen to lead to a
signal-to-noise-ratio (S/N) of at least 4 at $g$=20.2 mag and range
from 2160 $-$5500 sec. The central spectra have an instrumental
resolution of $R = \frac{\lambda}{\Delta \lambda} \sim 1800$ (i.e.,
FWHM $\sim$ 2.4 \AA~at 5000 \AA) and span 3800 $-$ 9200 \AA~ in
observed wavelength.  The SDSS spectroscopic pipeline performed the
basic reductions, including extraction, flux and wavelength
calibrations and removal of atmospheric bands. The spectra are shown
in Figure~\ref{SN_sloangals5_fig}, with and without stellar
contribution (see \S~\ref{pca_subsec} for more details).

\begin{figure}[!ht]
\includegraphics[scale=0.3,angle=90]{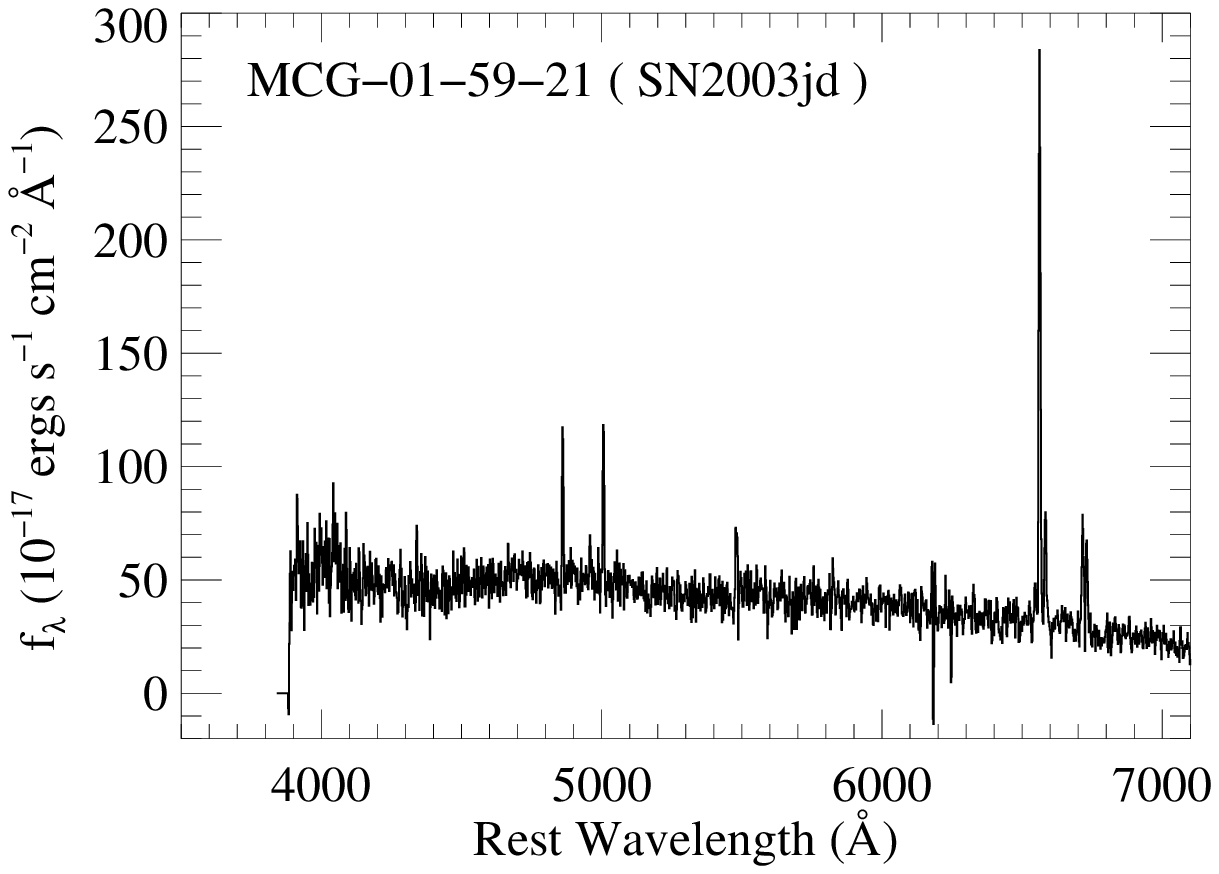} 
\caption{\singlespace Central host-galaxy spectra of SN~2003jd, taken from the 6dF
Galaxy Survey, de-redshifted using the value listed in
Table~\ref{sample_table}. See \S~\ref{obs_sec}.}
\label{SN03jd_nuc_fig}
\end{figure}

The central spectrum of SN~2003jd (Fig.~\ref{SN03jd_nuc_fig}) was
retrieved from the 6dF Galaxy Survey DR2\footnote{See
  http://www-wfau.roe.ac.uk/6dFGS/} \citep{jones04,jones05}. The
spectrum was acquired with the Six-Degree Field multi-object fiber
spectroscopy facility (with fibers 6$\farcs$7 in diameter) in two
observations using separate \emph{V} and \emph{R} gratings, that
combined result in an instrumental resolution of $R =
\frac{\lambda}{\Delta \lambda} \sim 1000$, range from 4000$-$ 7500
\AA, with typical S/N of $\sim$10 per pixel.


\subsection{Emission Line Measurements}\label{pca_subsec}

The focus of this paper is the analysis of the nebular \emph{emission}
lines of broad-lined SN Ic host galaxies. However, for spectra of
galaxy nuclei, the presence of stellar \emph{absorption} features may
contaminate the emission components, especially the Balmer lines
(e.g., \citealt{tremonti04,moustakas06}). This is not a concern for
spectra of HII regions near or at the position of the SN, as there the
light-weighted emission is contributed foremost by hotter stars with
weak Balmer lines. In our sample of \Nnucgalspec\ central spectra, we
encountered seven central spectra with significant stellar
features. To remove the starlight we utilize the method of stellar
template subtraction (see \citealt{ho04} for a full discussion of the
different techniques); in particular, we employ the standard principal
component analysis (PCA) technique developed by \citet{hao05} for use
on SDSS data (see also \citealt{greene04}). Using an approach
pioneered by \citet{bromley98} for the Las Campanas Redshift Survey
(see also \citealt{connolly95}), we employ the methods of \citet{hao05}
to use a library of SDSS spectra for absorption-line galaxies that is
transformed into an orthogonal basis of ``eigenspectra'' spanning the
variance in the sample. We model the absorption and continuum part of
the input galaxy spectrum as a linear combination of eight
eigenspectra plus an additional A-star component. The PCA method is
especially powerful since it does not assume single-metallicity
populations, uses the homogeneous survey-data of SDSS as templates,
need not include the galaxy velocity dispersion and spans a large
range in galaxy properties with a very small number of components. In
Figure~\ref{SN_sloangals5_fig} we plot the affected original spectra
(top) and spectra after subtracting the modeled stellar continuum
(bottom) : the SDSS spectra of \sndqhost, \snefhost, \snkshost,
\snqkhost, and \snihost, i.e., the host galaxies of supernovae 1997dq,
1997ef, 2005ks, 2006qk, and 2007I respectively. Thus, we arrive at
emission-line spectra.

For \snaphost, \sndahost, and \snfkhost, the host galaxies of
SN~2002ap, 2005da and 2005fk, respectively, we are unable to recover
any emission lines in their central regions, due to the low
surface-brightness or absence of \ion{H}{2} regions. Since
\snaphost~is a nearby, well-observed star-forming spiral galaxy, we
consult the literature for the emission-line fluxes of its \ion{H}{2}
regions as a function of radius (see \S~\ref{metal_subsec}). For the
hosts of SN~2005da and 2005fk, we are not able to measure
emission-line-based metallicities and refrain from using Lick indices,
as systematic offsets plague different metallicity methods. The low
surface-brightness or absence of \ion{H}{2} regions implies little
instantenous (i.e., in the last $\sim$ 3 million years) star formation
for these galaxies.

After properly correcting all spectra for their recession velocities
(as listed in Table~\ref{sample_table}), we measure optical emission
line properties by fitting Gaussians to the individual spectral lines
via the $splot$ routine in IRAF. The emission-line fluxes and their
statistical errors are given in Tables~\ref{lines_nuc_table} and \ref{lines_SNpos_table}. For the
derivation of the statistical errors, we follow \citet{perez03} and
use spectra before continuum-subtraction to measure their noise
properties. The total errors are computed by adding the statistical
errors and the 5\% flux calibration error in quadrature.

In the subsequent analysis, we correct the line fluxes for Galactic
reddening, according to \citet{schlegel98}, and for intrinsic
reddening by using the observed Balmer decrement, if H$\beta$ emission
line is detected. We assume case B recombination and thus, the
standard value of 2.86 as the intrinsic H$\alpha/H\beta$ ratio
\citep{osterbrock89}, and apply the standard Galactic reddening law
with $R_{V}=3.1$ \citep{cardelli89}. Tables \ref{prop_table_nuc} and
\ref{prop_table_atSNpos} list in column (9) the color excess due to the SN host galaxy $E(B-V)_{\rm{Host}}$
derived from the observed Balmer decrement, a robust indicator of the
amount of dust (e.g.,
\citealt{calzetti94,kewley02_red,moustakas06_intcomp}).

\section{Derived Host-Galaxy Properties}\label{hosts_sec}
\subsection{ Galaxy Luminosities}
We drew the host-galaxy luminosities from the Lyon-Meudon
Extragalactic Database (HyperLEDA)\footnote{See http://leda.univ-lyon1.fr/}
as their sample has been homogeneously compiled. We adopted as galaxy
luminosities the values listed as \emph{mabs} in HyperLEDA, i.e. the absolute
B-band magnitude, \emph{$M_B$}, corrected for Galactic and internal extinction and
k-corrections.

We applied a similar procedure to derive absolute magnitudes for the hosts of Sloan SN that were not listed in HyperLEDA. We retrieved the
host-galaxy Petrosian magnitudes  from SDSS DR5\footnote{See
http://cas.sdss.org/dr5/en/tools/explore/.} and corrected for Galactic
extinction using the tabulated reddening values and a standard
Galactic reddening law with $R_{V}=3.1$ \citep{cardelli89}. We applied
k-corrections calculated via \emph{kcorrect} (v.3.2\footnote{See
http://cosmo.nyu.edu/blanton/kcorrect/}), which was developed by
\citet{blanton03} for use with the SDSS filter set, in order to obtain
the absolute $B$-band magnitude \emph{$M_B$}. The full list is given in
column (2) of Tables \ref{prop_table_nuc} and
\ref{prop_table_atSNpos}.

\subsection{Metallicities and Star Formation Rates}\label{metal_subsec}

The nebular oxygen abundance is the canonical choice of metallicity
indicator for ISM studies, since oxygen is the most abundant metal in
the gas phase, only weakly depleted, and exhibits very strong nebular
lines in the optical wavelength range
\citep{tremonti04,kobulnicky04}. Thus, well-established diagnostic
techniques have been developed (e.g., \citealt{osterbrock89,pagel79}).
We compute host gas-phase oxygen abundance for both SN Ic (broad) and
of SN Ic (broad \& GRB) using three different and independent
metallicity diagnostics. Using three independent calibrations that
utilize different sets of lines for computing the oxygen abundance, we
can check if any of our results depend on the choice of
diagnostic. Furthermore, since different metallicity calibrations can
have large systematic offsets between each other
(\citealt{ellison05,ellison05_proc,kewley07}), we compute abundances
for host galaxies of both SN Ic (broad) and SN Ic (broad \& GRB) in
the same scale and compare their relative values. The three well-known
diagnostics used here are: 1) the iterative strong-line diagnostics
calibrated by \citet{kewley02} (henceforth KD02) and updated by
\citet{kobulnicky04}, 2) the calibration by \citet{mcgaugh91}
(henceforth M91) and 3) the diagnostic of \citet{pettini04} in the
direct electron temperature ($T_e$) scale. Details about each
calibration technique will be given below. For the SN Ic (broad), we
furthermore list central abundances in Table~\ref{prop_table_nuc} and
those at the SN position separately (Table~\ref{prop_table_atSNpos})
as spiral galaxies possess observed radial metallicity gradients, from
metal-rich centers to metal-poor outskirts (e.g.,
\citealt{vanzee98,bell00,kewley05}). For the SN observed with GRB, we
consulted the literature for the emission line fluxes at the GRB-SN
sites (Table~\ref{GRBSNsample_table}) and list the abundances computed
with the same three diagnostics in Table~\ref{GRBSN_prop_table}.

KD02 use a host of stellar population synthesis and photoionization
models to derive their metallicity diagnostics. Oxygen abundances
derived iteratively in this manner are listed in column (3) of
Tables~\ref{prop_table_nuc},
\ref{prop_table_atSNpos}, and \ref{GRBSN_prop_table}, corrected for extinction [column (9)].
For the metallicity range of log (O/H)+12$\geq$8.5, the KD02 method
mainly uses the \ntwo/\otwo~ratio, while for the low-metallicity
regime, the $R_{23}$ diagnostic, defined as $R_{23}$= (\otwo +
\othree)/H$\beta$, is usually employed.  If neither H$\beta$ nor
\otwo~are observed, we adopt the \ntwo/H$\alpha$ method from
\citet{kobulnicky04} with the KD02 calibration, assume a normal
ionization parameter ($q=2\times 10^7$) and note those cases in the
tables.

As alternatives, we also provide metallicity determinations using the
well-known strong-line calibration of \citet{mcgaugh91} (henceforth
M91) in column (4) and \citet{pettini04} (henceforth PP04) in column
(5). M91 provides a theoretical calibration of the $R_{23}$ parameter
via photoionization models that produce oxygen abundances that are
comparable in accuracy to direct methods that rely on the measurement
of nebular temperature. The PP04 calibration, on the other hand, is
predominantly based on an empirical fit to the electron temperature
(T$_e$)-abundances (using the lines of \othreeweak~and \otwo) of
\ion{H}{2} regions. This method is often referred to as the ``direct''
abundance method, as accurate abundance measurements can be obtained
directly from line ratios of auroral to nebular line intensities, such
as \othreeweak /$\lambda$5007 that indicate the electron
temperature. However, the $T_e$ method saturates at higher
metallicities (log (O/H)+12 $\geq$8.4), probably due to the additional
cooling of [O III] lines in the IR which is not accounted for in
current $T_e$ prescription, and hence PPO4 include six
$R_{23}$-derived metallicities to extrapolate to higher
metallicities. The PPO4 scale is typically $\sim$ 0.2 dex lower than
the theoretical strong line methods such as KD02. The exact cause of
this discrepancy is highly debated (for a discussion see e.g.,
\citealt{garnett04_cno,garnett04_m51,stasinska05}), and it might be
due to temperature variations that are not considered properly in the
T$_e$ prescription \citep{bresolin06}. For the host galaxies of SN Ic
(broad), we used the recommended [O III]/[N II]-based prescription of
PP04 (henceforth called PP04-O3N2). For the host galaxies of SN Ic
(broad \& GRB), we list in Table~\ref{GRBSN_prop_table} their oxygen
abundances in the $T_e$ scale, either via direct measurements of
\othreeweak\ or by converting KD02 abundances into $T_e$-based
ones. Since the PP04 calibration is tied to the $T_e$-abundances, the
inter comparison should be valid. We note that the GRB-SN abundances
in the scales of KD02 and $T_e$ were already computed in
\citet{kewley07}. We furthermore use the direct $T_e$-based abundance
for the host of \snaj\ as derived from the recent detection of
\othreeweak\ \citep{wiersema07}, which is the same as the
T$_e$-converted KD02 abundance from \citet{kewley07}.

We compute the uncertainties in the measured metallicities by
explicitly including the statistical uncertainties of the line flux
measurements and those in the derived SN host galaxy reddening, and by
properly propagating them into the metallicity determination. Details
are discussed in
\S~\ref{erroroxy_subsec}.

For \snaphost, the host of SN~2002ap, we took the measured emission
line fluxes of individual HII regions in M~74 and their deprojected
galactocentric radii from
\citet{mccall85,ferguson98,vanzee98,bresolin99} and
\citet{castellanos02}. Having computed the oxygen abundances in the
three metallicity scales, we fit for the observed M~74 abundance
gradient and obtain radial gradients of \gradkd $\pm$0.04
dex/$\rho_{25}$, \gradm $\pm$0.05 dex/$\rho_{25}$, and \gradpp
$\pm$0.05 dex/$\rho_{25}$,in the scales of the KD02, M91 and
PP04(O3N2), respectively, where $\rho_{25}$ stands for the isophotal
radius. With an extrapolated central metallicity as listed in
Table~\ref{prop_table_nuc} and the appropriate gradient, we compute
the metallicity at the radial de-projected distance of SN~2002ap
(4\arcmin38\arcsec), and list them in Table~\ref{prop_table_atSNpos}.
We note that the oxygen abundance at the SN position is similar to the
value quoted in \citet{smartt02} and \citet{crockett07} in the
appropriate scales. The uncertainty in the value of $\rho_{25}$ for
M~74 introduces the (small) uncertainty of 0.02 dex in the oxygen
abundance value, which we include in the uncertainty budget.

For \sndqhost~and \snbghost, the hosts of SN~1997dq and 2003bg, we do
not possess spectra taken at the SN positions, which are 52\arcsec and
30\arcsec, respectively, from the galaxy centers. At the distance of
SN~1997dq ($\sim$ 18 Mpc, from HyperLEDA) and of SN~2003bg ($\sim$ 16
Mpc, from HyperLEDA), the SN de-projected galactocentric distances
amount to $\sim$ 4.5 kpc and 2.4 kpc, respectively. Assuming usual
metallicity gradients of $\sim -$0.05 dex~kpc$^{-1}$ for $M_B=-20$ mag
spirals and of $\sim -$0.1 dex~kpc$^{-1}$ for $M_B=-17.5$ mag galaxies
\citep{garnett97,henry99}, i.e., similar to the host galaxies, we
adopt SN metallicity values that are 0.2 dex lower than that of the
nucleus for both galaxies (see Table~\ref{prop_table_atSNpos}). In
order to account for the uncertainty in the extrapolated abundances at
the SN site, we added in quadrature the spread in metallicity
gradients (choosing a conservative value of 0.1 dex) observed for
galaxies of corresponding luminosity to the uncertainty in the central
abundance.

We note that we were not able to compute reliable metallicities and
star formation rates for the host galaxy of SN~2005kz. Its spectrum is
dominated by the AGN and its line ratios of [N II]/H$\alpha$, [O
III]/H$\beta$, [O I]/H$\alpha$ are characteristic of a composite
starburst/Seyfert2 object \citep{kewley06_sdss}. Thus, we only list
SN~2005kz's host galaxy $M_B$ in Table~\ref{prop_table_nuc}, along
with the $M_B$ of hosts of SN~2005da and 2005fk.

Using the H$\alpha$ emission line strengths listed in
Tables~\ref{lines_nuc_table} and \ref{lines_SNpos_table}, we derive
extinction-corrected local H$\alpha$ luminosities and
extinction-corrected local star formation rates (SFRs)
\citep{kennicutt98}, and list them in columns (6) and (7) of
Tables~\ref{prop_table_nuc}, \ref{prop_table_atSNpos} and
\ref{GRBSN_prop_table}. We note that these values are lower limits to
the global SFR since the galaxies have larger angular diameters than
the slit sizes used, except for
\snkrhost, the host galaxy of SN~2005kr. When available, we use the
line ratio of
\stwo~and \stwol~combined with the Mappings photoionization code
\citep{sutherland93} and a 5-level model to compute electron densities
in the S+ zone (column 8).

\subsection{Uncertainties in Metallicity Measurements}\label{erroroxy_subsec}

We explicitly include the effect of the statistical errors in the line
measurements on the metallicity determinations. For each metallicity
diagnostic, we compute the minimum and maximum oxygen abundance value
for the given statistical uncertainties in the line measurements and
assign them as bounds for the most positive and negative errors. These
are listed in Tables~\ref{prop_table_nuc} and
\ref{prop_table_atSNpos}, directly following the reported abundance
value as determined from the measured line values. In the computation,
we also include the effect of the statistical error in the
H$\alpha$/H$\beta$ ratio that influences the measured Balmer decrement
and therefore the derived reddening. Because of logarithmic dependence
of the metallicity diagnostics on line ratios, the computed abundance
uncertainties are sometimes not symmetric around the reported values.
Furthermore, we checked whether a change in the Case B recombination
value from 2.85 to e.g., 3.1 (typical of AGN) makes a difference in
the derived metallicities from [N II]/[O II], and the effect is minor
($<$~0.03~dex).

In the optical, the choice of extinction curve makes
negligible difference to the extinction correction of optical line
ratios (see e.g., the review by \citealt{calzetti01}). We have tested this
ourselves by applying three different extinction curves to
[N II]/[O II]: the Whitford Reddening curve, as parameterized by
\citet{miller72}, the parameterization by \citet{cardelli89}, and that by \citet{osterbrock89}. 
We find that metallicities derived from [NII]/[OII] using different
optical extinction curves are identical. The same conclusion was
obtained by \citet{moustakas06} and \citet{moustakas06_intcomp},
who explored the effect of different extinction curve on oxygen
abundance determinations of a sample of 417 nearby galaxies with a
diverse range of galaxy types and histories.

For the SN-GRB sample, the line flux errors are not available for the
majority of GRB host spectra in the literature and we therefore assign
a uniform conservative error of 0.1 dex.

\section{Discussion}\label{results_sec}

We find that the host galaxies of broad-lined SN Ic have the following
properties: they are mid- to late-type spirals (consistent with the
host-galaxy morphological classifications found for the host galaxies
of SN Ib and SN Ic in \citealt{vandenbergh05} and references therein),
their $B$-band luminosities range between $-17 < M_B < -22$ mag, and
their central oxygen abundances span 8.63 $<$12+log(O/H)$_{\rm{KD02}}
<$9.15. Along with their SFRs and electron densities, they appear to
conform to the general population of normal local star forming
galaxies (for comparison with SDSS star forming galaxies, see
\ref{comp_subsec} and Figure~\ref{main_fig}).

\subsection{Comparison with SN-GRB Sample}\label{comp_subsec}

It is the main goal of this study to compare the physical
characteristics of host galaxies of SN Ic (broad) that had no observed
GRBs to those of SN-GRB hosts. 

First, we find that the SFR values are similar between the two
samples, with SFR(H$\alpha$)$\sim$ 0.1$-$3 \smy. One exception appears
to be \snbghost, the host of SN~2003bg, which has a value of $10^{-5}$
\smy, not corrected for extinction. However, most of our local SFR
values of SN Ic (broad) are lower limits to the global values, and
thus they are probably larger than those of the GRB-SN host
sample. Furthermore, considering the much smaller masses for the GRB
hosts as indicated by their lower luminosities, the $specific$ SFR
(SFR normalized to galaxy mass) of GRB hosts are larger than those of
SN Ic hosts, and of SDSS galaxies in general
\citep{christensen04,gorosabel05,sollerman05,savaglio06_conf,thoene06}.  A
comparison between the electron densities does not yield any
statistical significant results, as only one SN-GRB host has a
measured electron density. All measured electron densities lie within
the range of observed values for galaxies and HII regions.

\begin{figure*}[!ht]
\includegraphics[scale=0.7,angle=-90]{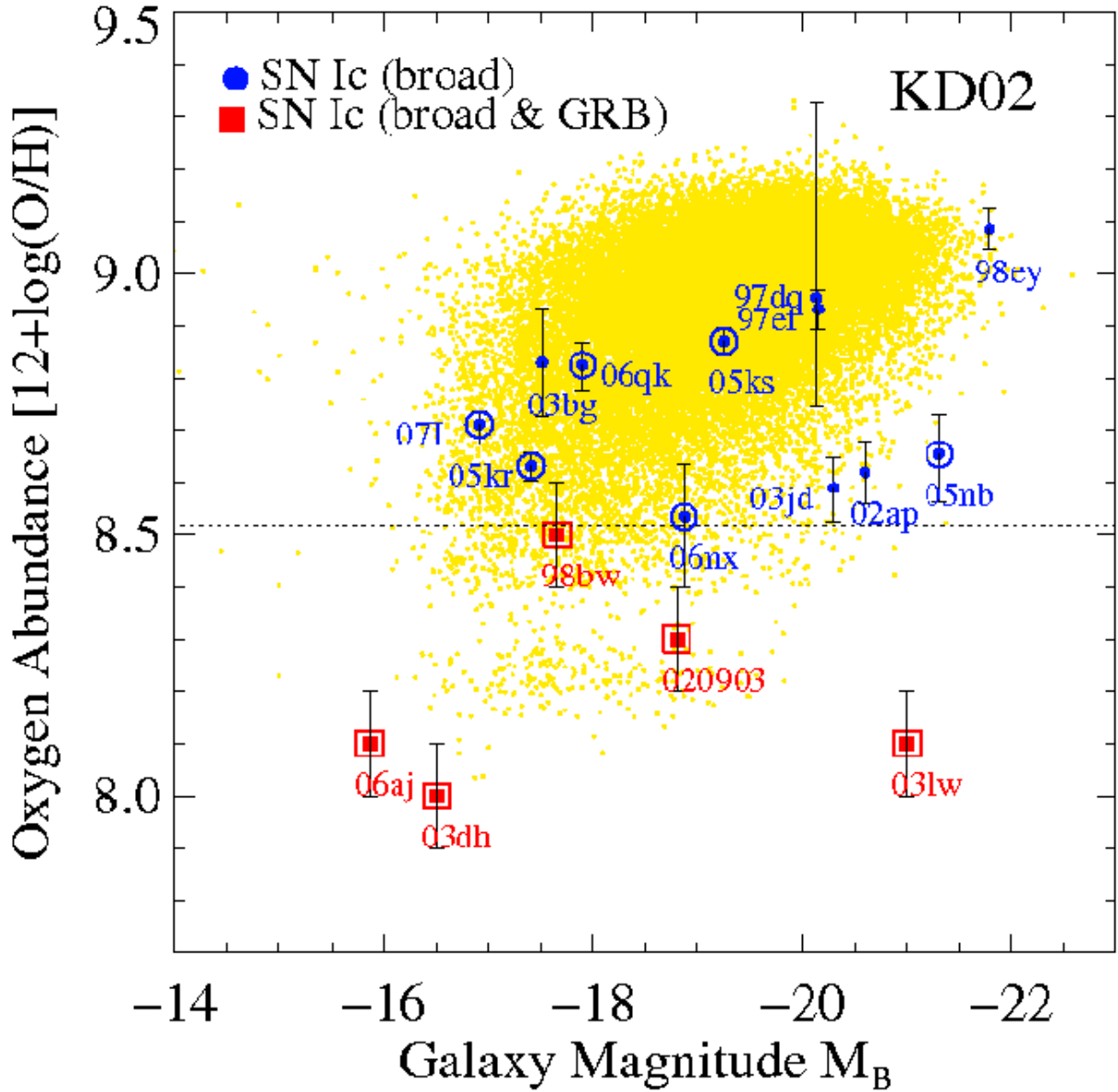}
\caption{\singlespace Host galaxy luminosity (\emph{$M_B$}) and host
  galaxy metallicity (in terms of oxygen abundance) at the sites of
  nearby broad-lined SN Ic (``SN Ic (broad)'': blue filled circles)
  and broad-lined SN Ic connected with GRBs (``SN (broad \& GRB)'':
  red filled squares; see also \citealt{stanek07,kewley07}). Extra
  circles and squares designate SN which were found in a non-targeted
  fashion. The oxygen abundances are in the \citet{kewley02} (KD02)
  scale and represent the abundance at the SN position. Due to radial
  metallicity gradients, the gas abundance at the SN position is lower
  than the central galaxy abundance for some SN (compare
  Tables~\ref{prop_table_nuc} and \ref{prop_table_atSNpos}). Labels
  represent the SN names while one (``020903'') refers to its
  associated GRB. Yellow points are values for local star-forming
  galaxies in SDSS (\citealt{tremonti04}), re-calculated in the
  \citet{kewley02} scale for consistency, and illustrate the empirical
  luminosity-metallicity ($L-Z$) relationship for galaxies. Host
  environments of GRBs are systematically less metal-rich than host
  environments of broad-lined SN Ic where no GRB was observed. The
  dotted line at 12+log(O/H)$_{\rm{KD02}}\sim$8.5 designates the
  apparent dividing line between SN with and without observed GRBs. }
\label{main_fig}
\end{figure*}

Next, Figure~\ref{main_fig} shows the main result of our comparison:
we plot host galaxy metallicity (as expressed in terms of oxygen
abundance 12+log(O/H) with the KD02 calibration) and host galaxy
luminosity (\emph{$M_B$}) of broad-lined SN Ic without observed GRBs
(called ``SN Ic (broad)'', circles) and with GRBs (called ``SN Ic
(broad \& GRB)'', squares). Objects whose host galaxies had not been
targeted during the discovery have an extra circle (for SN without
GRBs) or an extra square (for SN with GRBs) around their plotted
symbol. We note that three of the five broad-lined SN Ic found in the
lower luminosity galaxies ($M_B > -$19 mag) were discovered by the
SDSS-II SN Survey, which is a galaxy-impartial survey. For the
broad-lined SN Ic we only plot abundances measured at or extrapolated
to the SN position. These plotted values are lower than the values we
measure for the center of the same host galaxies (see
Table~\ref{prop_table_nuc}) presumably due to metallicity gradients as
discussed in \S~\label{metal_subsec}. The SN-GRB host abundances also
reflect the metallicity at the SN-GRB position, since they were either
measured specifically at the SN position (for \snbw), or the SN reside
in the nucleus of their dwarf-galaxy hosts that are chemically
homogeneous \citep{kobulnicky97}. Due to the short life times of the
massive SN progenitor stars ($t \la$ 10 Million years for $M_{ZAMS}
\ga$ 20 \sm, e.g., \citealt{woosley02}), we regard the metallicities
at the SN position as natal metallicities.

For reference, we also plot the luminosities of local (0.005$<z<$0.2)
SDSS star forming galaxies from \citet{tremonti04}. The plotted
central oxygen abundances were re-calculated in the KD02 scale, using
the published emission line fluxes, to be internally consistent. We
used the DR2 values with a SNR-cut of 5 for the emission line fluxes.
Their loci on Figure~\ref{main_fig} illustrate the well-studied
luminosity-metallicity ($L-Z$) relationship for disk galaxies (e.g.,
\citealt{rubin84,garnett02,tremonti04}), which has been also observed
for dwarf irregular galaxies
\citep{searle72,lequeux79,skillman89,pilyugin01,garnett02} and for
blue compact galaxies \citep{campos93,shi05}. We note that the SDSS
metallicities were derived using fibers with 3\arcsec in diameter,
corresponding to a mean projected fiber size of $\sim$4.6 kpc for that
sample \citep{tremonti04}, and that they therefore generally
reflect central metallicities (see also \citealt{kewley05}). Hence, a
detailed comparison between the distribution of abundances at the SN
sites and those of SDSS galaxies is not warranted, and the SDSS loci
should be regarded as indicating a general trend.

There is a bimodal distribution of host-galaxy luminosities and
metallicities: the luminosities of the hosts for the SN-GRB are
generally low, while the SN Ic (broad) hosts extend to high galaxy
luminosities. What is more, the SN-GRB have low metallicity
environments while SN Ic (broad) are found in systematically higher
metallicity environments. We plot as a dotted line the value of
12+log(O/H)$_{\rm{KD02}}\sim$8.5 that appears to be the dividing line
for the oxygen abundance between SN with and without observed GRBs.
While this GRB host-galaxy preference compared to other galaxies has
been noted before (e.g.,
\citealt{prochaska04,gorosabel05,sollerman05,modjaz06,stanek07,kewley07}),
we emphasize that we are comparing now the host galaxies of SN with
the same spectral characteristics. Thus, our comparison includes any
conditions required to produce envelope-stripped core-collapse SN and
we can test whether low metal abundance is a sufficient condition for
the evolution of the massive progenitor through the stages that lead
to a GRB jet. Using the host metallicities with the KD02 calibration
of SN Ic (broad \& GRB) and of SN Ic (broad) found in non-targeted
fashion, we conduct a simple Kolmogorov-Smirnov (K-S) Test. We find
that the probability that both sets of host metallicities have been
drawn from the same parent population is low, namely $\sim$ 3\%.

\begin{figure*}[!ht]
\hskip-0.5cm 
\includegraphics[scale=0.68,angle=-90]{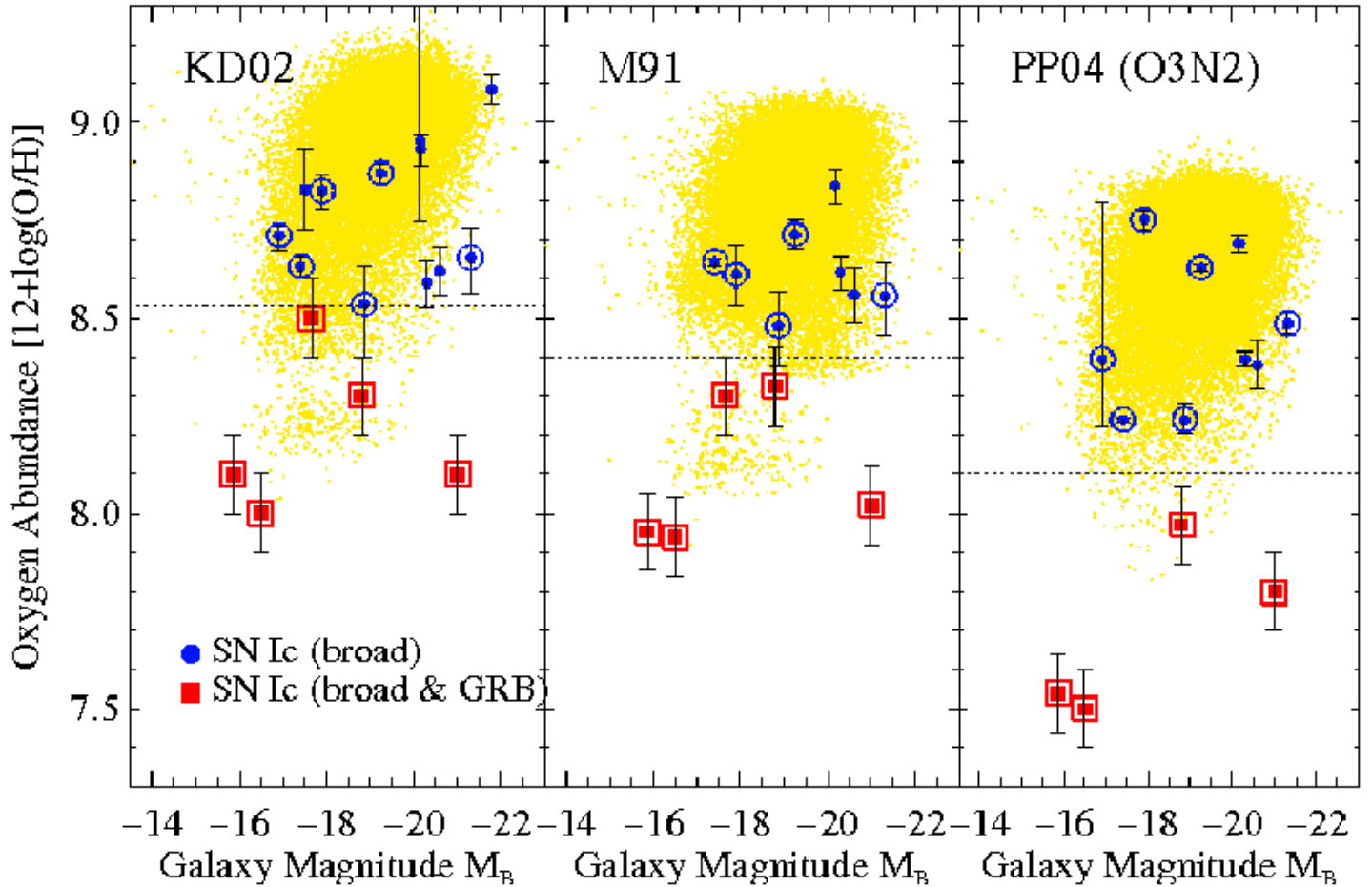}
\caption{\singlespace Similar to Figure~\ref{main_fig}, but using
  different metallicity diagnostics; \citet{kewley02} (KD02) as in
  Figure~\ref{main_fig} (\emph{left}); \citet{mcgaugh91} (M91,
  \emph{middle}); and \citet{pettini04} (PP04-O3N2), which is
  effectively on the $T_e$ scale, (\emph{right}). As before, host
  galaxy luminosity (\emph{$M_B$}) and host galaxy metallicity (in
  terms of oxygen abundance) at the SN site are plotted for nearby
  broad-lined SN Ic (``SN Ic (broad)''; blue filled circles) and for
  broad-lined SN Ic connected with GRBs (``SN (broad \& GRB)''; red
  filled squares). Extra circles and squares designate SN which were
  found in a non-targeted fashion. Host environments of GRBs are more
  metal-poor than environments of broad-lined SN Ic where no GRB was
  observed, independent of the abundance scale used. Yellow dots
  designate the SDSS galaxy values calculated in the respect
  metallicity scales.}
\label{Oxy3_fig}
\end{figure*}

Moreover, we plot in Figure~\ref{Oxy3_fig} the comparison between the
two host samples and the local SDSS galaxies in the metallicity scales
of M91 and of PP04-O3N2, with the latter being effectively in the
$T_e$ scale. While the absolute values of the abundances are different
in different scales, as expected, the bimodal distribution persists in
each scale and thus, is independent of the choice of metallicity
diagnostic. The K-S Test applied to the host abundances of SN Ic
(broad \& GRB) and SN Ic (broad) found in a non-targeted fashion
yields low probabilities of 4\% (M91-based abundances) and 3\%
($T_e$-based abundances) that they are drawn from the same
population. For each scale, we plot as a dotted line the boundary that
separates the two samples: 12+log(O/H)$_{\rm{M91}}\sim$8.4 and
12+log(O/H)$_{T_e}\sim$8.1. Although our sample is small our findings
are consistent with the hypothesis that low metal abundance is the
cause of some very massive stars becoming SN-GRB.

Of course, a question of interest for GRB-modelers is how these
cut-off values in oxygen abundance compare to solar metallicity. The
answer is not clear due to the problems that plague absolute
metallicity determination: it depends on which of the three
metallicity scales is adopted and whether one uses the high value for
solar oxygen abundance (12 + log(O/H)=8.9, \citealt{delahaye06}) or
the lower, revised solar abundance (12 + log(O/H)=8.7,
\citealt{asplund05}). Thus, the cut-off value can range between 0.2
$-$0.6 \sz. Nevertheless, our main conclusions do not depend on the
absolute metallicity scale.

We note that \citet{tremonti04} computed the metallicity of the host
galaxy of SN~2007I with their diagnostics, using the same SDSS
emission line spectrum we used. Their value,
12+log(O/H)$_{\rm{Tremonti 04}}$ = 8.3, is 0.4 dex lower than
12+log(O/H)$_{\rm{KD02}}$ = 8.71 $\pm$ 0.04, but similar to
12+log(O/H)$_{\rm{PP04 03N2}}$ = 8.39$^{+ 0.4}_{-0.17}$. We attribute
this difference to the absence of [\ion{O}{2}] in the galaxy spectrum
which precludes the usage of the reliable KD02 methods (using
[NII]/[OII] or $R_{23}$), and instead necessitates using
[NII]/H$\alpha$. Thus, the reliable metallicity for this galaxy is in
the PP04 scale. Nevertheless, this caveat does not change our
conclusions as the GRB-SN metallicities are even lower than
12+log(O/H)$_{\rm{PP04 03N2}}$ = 8.4.

At face value, our results differ from various studies in the
literature, which have concluded that high-$z$ GRB hosts are not
necessarily low-metallicity systems. However, there might be intrinsic
differences in GRB population at low and high $z$ $-$ as mentioned, the
low-$z$ GRBs might constitute a different class of low-luminosity GRBs
that are not detected at higher $z$ \citep{cobb06,liang06,guetta06}.
Moreover, most of the techniques for measuring abundances at higher
$z$ are different from our direct approach: they either use absorption
techniques in GRB afterglows (\citealt{prochaska07} and references
therein) or infer the metallicity using galaxy luminosity and color
via the $L-Z$ relation \citep{berger06}. Attempts to directly
measure GRB host metallicities at higher redshifts via emission line
diagnostics remain inconclusive (\citealt{savaglio06_conf}, S. Savaglio in
prep.).

Recently, \citet{wolf07} suggested that if there were a sharp cut-off
for the metallicity dependence of cosmological GRBs, assuming that the
GRB-hosts follow the standard $L-Z$ relationship at $z\sim0.7$, it
would lie at around 12 + log(O/H)$_{\rm{lim}} \sim$ 8.7 $\pm$ 0.3 in
the scale of \citet{kobulnicky04}. They claimed to rule out GRB models
that require sharp metallicity cut-offs well below one-half the solar
abundance (assuming the value of 8.7 as the solar oxygen abundance),
since those models would predict GRB hosts that need to be much less
luminous than the observed hosts in the sample of
\citet{fruchter06}. Interestingly, even though their assumptions rely
on the conforming behavior of GRB hosts, and they investigate
cosmological GRBs, their findings are consistent with our
results. Ongoing attempts to directly measure host galaxy
metallicities of cosmological GRBs (e.g., \citet{savaglio06_conf})
are expected to clarify this debate, though long integration times and
NIR spectra will be required for the bulk of the cosmological GRBs due
to their high redshift (with a mean $z \sim$ 3 for the Swift GRBs).

\subsection{Viewing Angle Effects and Caveats}\label{caveats_subsec}

Cosmological GRB are highly collimated events, with beaming angles
$\theta \sim 4 - 10\deg$ \citep{frail01,guetta05}, while the
low-luminosity, nearby GRB appear to be less beamed
\citep{guetta06}. Thus, one might argue that the lack of observed
gamma-ray emission for the SN Ic (broad) presented in this study is
due to viewing angle effects, i.e. the GRB occurred but our
line-of-sight was not aligned with the jet axis.

We have good reason to believe that this is probably not the case for
at least six of the twelve of the objects in our sample.
\citet{soderberg06_radioobs} searched for radio emission in 68 local
SN Ib and SN Ic at late times (after 0.5$-$20 years), when the GRB, if
present, is expected to become isotropic and emit in the radio
regime. They did not detect late-time radio emission and therefore
exclude an off-axis jet-driven SN engine in the following SN of our
sample: SN~1997dq, SN~1997ef, SN~1998ey, SN~2002ap, SN~2003jd, and in
addition, SN~2005da \citep{soderberg05_05da} and SN~2005kz
\citep{soderberg05_05kz}. We note that the conclusions of
\citet{soderberg06_radioobs} rely on modeling GRB radio properties
based on those observed for \snbw. Including \snaj,
\citet{soderberg06_06aj} infer that less than one in three broad-lined
SN may have a radio luminosity comparable to \snaj. For SN~2003bg,
strong early-time X-ray and radio emission was detected
\citep{soderberg03_03bg,pooley03} and interpreted as signs of
interactions between SN ejecta and dense CSM \citep{soderberg06_03bg}.

There was no such radio search for emission in the SN found by
the Sloan-II SN survey from our sample. Thus, the viewing angle issue
remains a caveat for these SN.

As noted above, \citet{mazzali05_03jd} argue for an off-axis jet in
SN~2003jd, since they interpret double-peaked oxygen and magnesium
profiles observed in its late-time spectra as signs of an aspherical,
axisymmetric explosion caused by a jet.  However, asphericities have
been observed in other core-collapse SN (e.g, \citealt{leibundgut91}),
most notably in SN~1987A (e.g., \citealt{wang02}), and might not
necessarily signal the presence of a relativistic jet of
material. Polarization studies suggest clear asphericity in ejecta of
all types of core-collapse SN, which increases with decreasing
envelope mass (see \citealt{leonard05} for a review). These results
indicate that asphericity might be an ubiquitous feature of the
conventional neutrino-driven process of core collapse, occasionally
damped by the additional envelopes of hydrogen and helium. In order to
fully observationally probe the range of geometries realized during
the explosion of massive stars, a systematic study of late-time
spectral features for a large sample of stripped-envelope
core-collapse supernovae (i.e., SN Ib, Ic, IIb) is necessary. Such a
study is presently underway and includes a SN Ib, SN~2004ao that
displays a double-peaked profile of oxygen with the same width and
characteristics as seen in SN~2003jd \citep{modjaz_thesis}.

Furthermore, \citet{li06_grbsn} describes tentative evidence for a
relationship between $E_{pk}$, the peak of the gamma-ray spectrum of
four GRBs with associated SN, and the peak luminosity ($M_{SN}$) of
their associated SN, such that GRBs with harder gamma-ray spectra have
associated SN that are more luminous. Furthermore, he suggests that the
$E_{pk}$ of the putative GRBs if they occurred in association with
supernovae 2002ap and 1997ef would have been located in the UV regime
due to the low SN luminosities and that those GRBs would have had very
low gamma-ray energies \citep{li06_grbsn}. If true, those explosions
would hardly qualify as bona fide ``gamma-ray'' bursts. In any case,
caution should be exercised when extending a relationship based on
four objects to a range outside that probed by the data, and future
nearby GRB-SN ought to be used to verify the $E_{pk}$-$M_{SN}$
correlation of \citet{li06_grbsn}.

Thus, we conclude that observations do not support associated off-axis
GRBs with six broad-lined SN Ic, while we cannot exclude the
possibility of off-axis GRBs in the SDSS-SN of our sample.

\section{Conclusions and Summary}\label{summary_sec}

In this paper we presented spectroscopic data of a statistically
significant set of host galaxies of \NblSNIc~nearby ($z <
$\hightestzSNIc) broad-lined SN Ic with no observed GRBs. Using the
galaxy emission-line measurements corrected for stellar absorption and
extinction, we derived central oxygen abundances and abundances at the
SN position based on strong-line diagnostics that span
$8.5<$12+log(O/H)$_{\rm{KD02}}<$9.1 on the \citet{kewley02} scale,
$8.5<$12+log(O/H)$_{\rm{M91}}<$8.9 on the \citet{mcgaugh91} scale, and
$8.2<$12+log(O/H)$_{PP04}<$8.9 on the \citet{pettini04} scale, which
effectively uses the $T_e$ scale. Furthermore, we computed local
star formation rates for central regions, as well as for regions at
the position of the SN, and drew from the literature the host galaxy
B-band luminosities, which range between $-17 < M_B < -22$ mag. This
study, which constructed local chemical abundances for the locations
of individual supernovae of type Ic (broad), is the first of its
kind.  To our knowledge, a broader application of this approach to
other core-collapse supernovae has not been carried out. 

We compared the properties of our host sample with the properties of
five nearby SN-GRB hosts, for which we derived chemical abundances
using the same three metallicity diagnostics as for SN without
observed GRBs. Broad-lined SN Ic without GRBs tend to consistently
inhabit more metal-rich environments, and their host galaxies, for the
$same$ luminosity range ($-17 < M_B < -21$ mag), are systematically
more metal-rich than corresponding GRB host galaxies. The trend is
independent of the choice of diagnostic and cannot be due to selection
effects as we include six SN found in a similar non-targeted manner as
GRB-SN. The boundary between broad-lined SN Ic that have a GRB
accompanying them and broad-lined SN Ic without a GRB lies at an
oxygen abundance of $\sim$ 12+log(O/H)$_{\rm{KD02}}\sim$ 8.5, which
corresponds to 0.2$-$0.6 \sz\ depending on the adopted metallicity
scale and solar abundance value. This correlation is consistent with
the argument that (local) long-duration GRBs require low-metallicity
environments for jet production. If this correlation holds up as the
samples grow, it would be consistent with the idea that low metal
abundance is a significant factor that allows a very massive star to
become both a broad-lined SN Ic and a GRB when it collapses. On the
other hand, if broad-lined SN Ic without an accompanying GRB are found
in galaxies with low metal abundance, then a more complicated set of
conditions than simply metal abundance will need to be considered.

Since 3/5 of the broad-lined SN Ic harbored in low-luminosity galaxies
were found by the SDSS-II SN Survey, we note the importance of
rolling, galaxy-impartial SN surveys to uncover the population of SN
in dwarf galaxies. Future impartial, large and deep photometric
surveys such as Pan-STARRS \citep{tonry05}, Skymapper
\citep{granlund06}, and the Large Synoptic Survey Telescope (LSST)
\citep{pinto04} will hence greatly contribute to the understanding and
study of stellar explosions in different galactic environments.
Finally, we recommend observational metallicity studies of a
statistically significant set of SN Ib, SN Ic, broad-lined SN Ic and
SN II as to investigate any metallicity effects on the properties of
core-collapse SN and on the extent of prior mass loss, which are
expected on theoretical (e.g., \citealt{woosley02,woosleyjanka05}) and
stellar evolutionary grounds \citep{crowther06}.

\section*{Acknowledgments}

M.M. would like to thank Jesse Leaman for making data available before
publication and John Moustakas for compiling the published HII region
line fluxes for M~74. We also acknowledge Christy Tremonti for making
available the SDSS galaxy parameters in a convenient
form. M. M. acknowledges helpful discussions with M. Geller, W. Brown,
K. Rines, C. Tremonti, C. Stubbs, R. Narayan, N. Smith and
D. Sasselov. We thank the service observers of the 1.5m FLWO for
taking some of the spectra presented here.

Observations reported here were obtained at the MMT Observatory, a
joint facility of the Smithsonian Institution and the University of
Arizona, and at the F.L Whipple Observatory, which is operated by the
Smithsonian Astrophysical Observatory. Supernova research at Harvard
University has been supported in part by the National Science
Foundatation grant AST06-06772 and R.P.K. in part by the Kavli
Institute for Theoretical Physics NSF grant PHY99-07949. Lisa Kewley and
Jenny Greene are supported by a Hubble Fellowship.

Furthermore, this research has made use of NASA's Astrophysics Data
System Bibliographic Services (ADS), the HyperLEDA database and the
NASA/IPAC Extragalactic Database (NED) which is operated by the Jet
Propulsion Laboratory, California Institute of Technology, under
contract with the National Aeronautics and Space Administration.

Funding for the SDSS and SDSS-II has been provided by the Alfred
P. Sloan Foundation, the Participating Institutions, the National
Science Foundation, the U.S. Department of Energy, the National
Aeronautics and Space Administration, the Japanese Monbukagakusho, the
Max Planck Society, and the Higher Education Funding Council for
England.

The SDSS is managed by the Astrophysical Research Consortium for the
Participating Institutions. The Participating Institutions are the
American Museum of Natural History, Astrophysical Institute Potsdam,
University of Basel, Cambridge University, Case Western Reserve
University, University of Chicago, Drexel University, Fermilab, the
Institute for Advanced Study, the Japan Participation Group, Johns
Hopkins University, the Joint Institute for Nuclear Astrophysics, the
Kavli Institute for Particle Astrophysics and Cosmology, the Korean
Scientist Group, the Chinese Academy of Sciences (LAMOST), Los Alamos
National Laboratory, the Max-Planck-Institute for Astronomy (MPIA),
the Max-Planck-Institute for Astrophysics (MPA), New Mexico State
University, Ohio State University, University of Pittsburgh,
University of Portsmouth, Princeton University, the United States
Naval Observatory, and the University of Washington.

\bibliographystyle{apj}
\bibliography{refs}

\clearpage

\begin{deluxetable}{lcccl}
\tabletypesize{\scriptsize}
\singlespace
\tablewidth{0pt}
\tablecaption{ Sample of Broad-lined SN Ic Associated with GRBs}
\tablehead{\colhead{GRB/SN Name} & 
\colhead{Host Galaxy} & 
\colhead{Hubble Type} & 
\colhead{Redshift $z$} &
\colhead{Refs} \\
\colhead{} &
\colhead{} &
\colhead{} &
\colhead{} &
\colhead{} }
\startdata 
\snbw &  ESO 184-G82            &         SB &     0.0085 & 1, 2     	      \\
\snlw &  Anon.			& \nodata    & 	   0.1055 & 3, 4      	 \\
XRF 020903 &  Anon.			& \nodata    &	   0.2500 & 2, 5 	   \\
\snaj &  Anon.			& \nodata    &	   0.0335 & 6, 7	    \\
\sndh &  Anon.			& \nodata    & 	   0.1685 & 1, 8, 9  	      
  
\enddata

\tablerefs{(1) \citet{sollerman05}; (2) \citet{hammer06}; (3) \citet{prochaska04}; (4) \citet{margutti07}; (5) \citet{bersier06}; 
(6) \citet{modjaz06}; (7) \citet{wiersema07}; (8) \citet{gorosabel05}; (9) \citet{thoene06}.}
\label{GRBSNsample_table}
\end{deluxetable}

\begin{deluxetable}{lcccccccl}
\tabletypesize{\scriptsize}
\singlespace
\tablewidth{0pt}
\tablecaption{ Sample of Broad-lined SN Ic Without Observed GRB}
\tablehead{\colhead{SN Name} & 
\colhead{Host Galaxy} & 
\colhead{Hubble Type} & 
\colhead{Redshift $z$} &
\colhead{RA Offset} &
\colhead{DEC Offset} &
\colhead{Refs} &
\colhead{Disc Type\tablenotemark{a}} \\  
\colhead{} &
\colhead{} &
\colhead{} &
\colhead{} &
\colhead{[~\arcsec~]\tablenotemark{b}} &
\colhead{[~\arcsec~]\tablenotemark{b}} &
\colhead{} &
\colhead{} &
\colhead{} }
\startdata
SN 1997dq &                   NGC3810 &         SA(rs)c &      0.0033 &   43.W &   29.N & 1, 2  & T       \\
SN 1997ef &                   UGC4107 &         SA(rs)c &      0.0117 &   10.E &   20.S & 3, 4  & T      \\
SN 1998ey &                   NGC7080 &          SB(r)b &      0.0161 &   18.W &   20.N & 5, 6  & T       \\
SN 2002ap &                       M74 &          SA(s)c &      0.0022 &  258.W &  108.S & 7, 8   & T    \\
SN 2003bg &              MCG-05-10-15 &          SB(s)c &      0.0044 &   16.W &   25.S & 9, 10  & T      \\
SN 2003jd &              MCG-01-59-21 &         SAB(s)m &      0.0188 &    8.E &    8.S & 11, 12 & T      \\
SN 2005da &                  UGC11301 &              Sc &      0.0150 &  108.W &   35.N & 13, 14 & T       \\
SN 2005fk &       J211520.09-002300.3 &          \ldots &     0.2650 &    4.W &    2.N &  15   & Non-T   \\
SN 2005nb &                  UGC07230 &      SB(s)d-pec &     0.0238 &    1.5W &   5.N &  16   & Non-T   \\ 
SN 2005kr &       J030829.66+005320.1 &          \ldots &     0.1345 &   0.0E &   0.1N &  17   & Non-T   \\
SN 2005ks &       J213756.52-000157.7 &          \ldots &     0.0987 &   0.0E &   0.8N &  17   & Non-T   \\
SN 2005kz &             MCG+08-34-032 &        Sc;LINER &     0.0270 &   16.W &    10.N &  18, 19 & T    \\ 
SN 2006nx &       J033330.43-004038.0 &          \ldots &     0.1370 &   0.2E &   0.2S &  20  & Non-T    \\
SN 2006qk &       J222532.38+000914.9 &          \ldots &     0.0584 &   0.0E &   0.2N &  21  & Non-T    \\
SN 2007I &       J115913.13-013616.1 &          \ldots &      0.0216 &   0.8E &   0.8S &  22, 23  & Non-T   
\enddata
\tablenotetext{a}{Discovery Type: T = SN host galaxy was targeted; Non-T = SN host galaxy was not targeted. }
\tablenotetext{b}{Offset from the center of host galaxy as listed in discovery IAUCs or as derived by comparing SN and host galaxy coordinates.}

\tablerefs{(1) \citet{nakano97_97dqdisc}; (2) \citet{matheson01};(3)
\citet{hu97_97efdisc}; (4) \citet{hu97_97efid}; (5)
\citet{schwartz98_98eydisc}; (6) \citet{garnavich98_98eyid}; (7)
\citet{nakano02_02apdisc}; (8) \citet{kinugasa_02apid}; (9)
\citet{chassagne_03bgdisc}; (10) \citet{filippenko03_03bg}; (11)
\citet{burket_03jddisc}; (12) \citet{filippenko_03jdid}; (13)
\citet{lee05_05dadisc}; (14) \citet{modjaz05_05daid}; (15)
\citet{barentine_05fk}; (16) \citet{quimby06_05nb}; (17)
\citet{barentine_05kr} ; (18) \citet{puckett05_05kzdisc}; (19)
\citet{filippenko05_05kzid}; (20) \citet{bassett_06nx}; (21)
\citet{bassett_06qk}; (22) \citet{lee07_07Idisc}; (23)
\citet{blondin07_07Iid}.  }
\label{sample_table}
\end{deluxetable}

\begin{deluxetable}{lcccccccccccl} 
\singlespace
\tablewidth{0pt}
\tabletypesize{\scriptsize}
\tablecaption{Journal of Spectroscopic Observations of Host Galaxies of Broad-lined SN Ic }
\tablehead{\colhead{SN} &
\colhead{UT Date} &
\colhead{Julian Day} &
\colhead{Telescope\tablenotemark{a}} &
\colhead{Observed Range}  &
\colhead{Resolution} &
\colhead{Airmass\tablenotemark{b}} & 
\colhead{Slit} &
\colhead{Exp.} &  
\colhead{Spectrum-type\tablenotemark{c}} \\  
\colhead{} &
\colhead{} &
\colhead{- 2,450,000} &
\colhead{} &
\colhead{(\AA)} &
\colhead{(\AA)} &
\colhead{} &
\colhead{($^{\prime\prime}$)} &
\colhead{(sec)} & 
\colhead{ } }
\startdata

        SN~1997ef & 1997-11-26.39 & 778.89 & FLWO & 3720-7540 & 7.0 & 1.12 & 3.00 & 1200 & SN \\
	SN~1997ef & 1997-11-28.52 & 781.02 & FLWO & 3720-7540 & 7.0 & 1.11 & 3.00 & 1200 & SN \\ 
	SN~1997ef & 1997-11-29.42 & 781.92 & FLWO & 3720-7540 & 7.0 & 1.06 & 3.00 & 1200 & SN \\
	SN~1997ef & 1997-11-30.34 & 782.84 & FLWO & 3720-7540 & 7.0 & 1.19 & 3.00 & 1200 & SN~ \\ 
	SN~1997ef & 1997-12-01.33 & 783.83 & FLWO & 3720-7540 & 7.0 & 1.22 & 3.00 & 1200 & SN~ \\ 
	SN~1997ef & 1997-12-04.34 & 786.84 & FLWO & 3720-7540 & 7.0 & 1.17 & 3.00 & 1200 & SN~ \\ 
	SN~1997ef & 1997-12-05.31 & 787.81 & FLWO & 3720-7540 & 7.0 & 1.26 & 3.00 & 1200 & SN~ \\ 
	SN~1997ef & 1997-12-06.33 & 788.83 & FLWO & 3720-7540 & 7.0 & 1.17 & 3.00 & 1200 & SN~ \\ 
	SN~1998ey & 1998-12-12.14 & 1159.64 & FLWO & 3720-7540 & 7.0 & 1.52 & 3.00 & 1200 & SN~ \\ 
	SN~2002ap & 2006-11-20.23 & 4059.73 & FLWO & 3479-7417 & 7.0 & 1.04 & 3.00 & 600 & Central \\ 
	SN~2003bg & 2006-09-21.0 & 3999.50 & LCO & 3800-10000 & 7.0 & 1.00 & 1.25 & 5$\times$600 & Central \\ 
	SN~2003jd & 2003-12-19.13 & 2992.63 & MMT & 3650-8850 & 8.0 & 1.49 & 1.00 & 900 & SN \\ 
	SN~2003jd & 2003-12-20.7 & 2993.57 & MMT & 3620-8830 & 8.0 & 1.28 & 1.00 &  3$\times$1200 & SN \\ 
	SN~2005da & 2006-10-25.08 & 4033.58 & MMT & 3340-8521 & 8.0 & 1.25 & 2.00 & 600 & Central \\ 
	SN~2005fk & 2006-09-21.0 & 3999.50 & LCO & 3800-10000 & 7.0 & 1.21 & 3.00 & 4$\times$1800 & Central \\
       	SN~2005kr & 2006-09-21.0 & 3999.50 & LCO & 3800-10000 & 7.0 &  1.38 & 1.25 & 3$\times$1800 & SN, Central \\ 
	SN~2005kz & 2006-10-25.11 & 4033.61 & MMT & 3340-8521 & 8.0 & 1.19 & 2.00 & 600 & Central \\ 
	SN~2005nb & 2006-01-09.51 & 3745.01 & FLWO & 3500-7410 & 7.0 & 1.04 & 3.00 & 1800 & SN \\ 
	SN~2006nx & 2006-12-13.0 & 4082.50 & LCO & 4060-10600 & 7.0 & 1.40 & 1.00 & 2$\times$1200 & SN, Central
\enddata
\tablenotetext{a}{FLWO = Tillinghast 1.5m/FAST; LCO = 6.5 m Clay Telescope of the Magellan Observatory located at Las Campanas Observatory/LDSS-3; MMT = 6.5m MMT/Bluechannel.\\}
\tablenotetext{b}{Airmass at the middle of the set of observations.\\}
\tablenotetext{c}{SN = Spectrum at the position of the SN; Central = Spectrum of the central part of the host galaxy. \\}
\label{obs_table}
\end{deluxetable}
\clearpage


\begin{deluxetable}{lccccccccl}
\singlespace
\tablewidth{0pt}
\tabletypesize{\scriptsize}
\tablecaption{Central Galaxy Emission Line Measurements}
\tablehead{\colhead{Emission Line} & 
\colhead{SN~1997dq} & 
\colhead{SN~1997ef } &
\colhead{SN~2003bg} &
\colhead{SN~2003jd} &
\colhead{SN~2005kr\tablenotemark{a}} &
\colhead{SN~2005ks\tablenotemark{a}} &
\colhead{SN~2006nx\tablenotemark{a}} &
\colhead{SN~2006qk\tablenotemark{a}} &
\colhead{SN~2007I\tablenotemark{a}}
}
\startdata
$[$OII]~$\lambda \lambda 3726,29$  & \ldots & \ldots & \ldots & \ldots & 13.7 $\pm$ 1.4 & 170 $\pm$ 19 & 15.2 $\pm$ 2.5 & 102 $\pm$ 14 & \ldots  \\
H$\delta$~$\lambda 4101$  & \ldots  & \ldots  & \ldots  & \ldots  & 1.23 $\pm$ 0.16 & \ldots  & \ldots  & \ldots  & \ldots   \\
H$\gamma$~$\lambda 4340$  & \ldots  & 67.3 $\pm$ 8.4 & \ldots  & 1421 $\pm$ 330 & 3.04 $\pm$ 0.32 & \ldots  & \ldots  & 15.7 $\pm$ 3.4 & \ldots   \\
H$\beta$~$\lambda 4861$  & 68 $\pm$ 10 & 239 $\pm$ 25 & \ldots  & 4620 $\pm$ 530 & 8.20 $\pm$ 0.83 & 69.6 $\pm$ 7.3 & 6.9 $\pm$ 1.1 & 43.7 $\pm$ 5.0 & 28.7 $\pm$ 2.9  \\
$[$OIII]~$\lambda 4959$  & \ldots  & \ldots  & \ldots  & 940 $\pm$ 250 & 8.40 $\pm$ 0.85 & 13.7 $\pm$ 2.0 & 6.87 $\pm$ 0.94 & \ldots  & 11.9 $\pm$ 2.9 \\
$[$OIII]~$\lambda 5007$  & \ldots  & 28.5 $\pm$ 5.7 & \ldots  & 4300 $\pm$ 630 & 24.3 $\pm$ 2.4 & 51.2 $\pm$ 5.5 & 20.9 $\pm$ 2.2 & 15.3 $\pm$ 2.6 & 57.2 $\pm$ 7.1  \\
HeI~$\lambda 5876$  & \ldots  & \ldots  & \ldots  & \ldots  & 1.15 $\pm$ 0.14 & \ldots  & \ldots  & \ldots  & \ldots   \\
$[$OI]~$\lambda 6300$  & \ldots  & 20.5 $\pm$ 3.5 & \ldots  & 720 $\pm$ 310 & 1.03 $\pm$ 0.13 & 9.5 $\pm$ 1.3 & \ldots  & \ldots  & \ldots  \\
$[$NII]~$\lambda 6548$  & 90 $\pm$ 40 & 131 $\pm$ 13 & \ldots  & 1400 $\pm$ 300 & 0.72 $\pm$ 0.11 & 34.6 $\pm$ 3.6 & \ldots  & 37.2 $\pm$ 4.1 & \ldots  \\
H$\alpha$~$\lambda 6563$  & 496 $\pm$ 97 & 1150 $\pm$ 110 & 2.35 $\pm$ 0.26 & 19000 $\pm$ 2000 & 28.4 $\pm$ 2.8 & 272 $\pm$ 27 & 33.1 $\pm$ 3.4 & 230 $\pm$ 23 & 119 $\pm$ 13 \\
$[$NII]~$\lambda 6584$  & 280 $\pm$ 64 & 407 $\pm$ 41 & 1.02 $\pm$ 0.14 & 4370 $\pm$ 550 & 2.41 $\pm$ 0.26 & 92.3 $\pm$ 9.3 & 2.73 $\pm$ 0.51 & 86.2 $\pm$ 8.8 & 20.0 $\pm$ 3.2 \\
$[$SII]~$\lambda 6717$  & 114 $\pm$ 28 & 149 $\pm$ 16 & \ldots  & 4250 $\pm$ 490 & 4.40 $\pm$ 0.45 & 65.6 $\pm$ 6.8 & \ldots  & 59.8 $\pm$ 6.8 & 27.4 $\pm$ 3.7 \\
$[$SII]~$\lambda 6731$  & 62 $\pm$ 19 & 111 $\pm$ 11  & \ldots  & 3210 $\pm$ 390 & 2.77 $\pm$ 0.29 & 48.4 $\pm$ 5.1 & \ldots  & 35.9 $\pm$ 4.6 & 23.6 $\pm$ 3.4 

\enddata
\tablecomments{Rest emission-line fluxes in units of $10^{-17}$
\ergscm~before extinction correction. Errors include statistical
measurement uncertainties. Column headings designate the SN names
whose host-galaxy line fluxes are presented; see
Table~\ref{sample_table} for corresponding galaxy names. We do not
list fluxes the hosts of SN~2002ap, 2005da and 2005fk, since no emission lines
were detected in the galaxy spectra.\\}
\tablenotetext{a}{Central spectra represent also spectra taken at the
SN position since the SN occurred in the galaxy nucleus. See
Table~\ref{sample_table}. \\ }
\label{lines_nuc_table}
\end{deluxetable}


\begin{deluxetable}{lcccccccl}
\tablewidth{0pt}
\singlespace
\tabletypesize{\scriptsize}
\tablecaption{ Galaxy Emission Line Measurements at SN Position if Different from Galaxy Center }
\tablehead{\colhead{Emission Line} & 
\colhead{SN~1997ef} & 
\colhead{SN~1998ey } &
\colhead{SN~2003jd} &
\colhead{SN~2005kz} &
\colhead{SN~2005nb} 
}
\startdata
$[$OII]~$\lambda \lambda 3726,29$  & 960 $\pm$ 100 & \ldots & 717 $\pm$ 75 & 334 $\pm$ 35 & 1170 $\pm$ 130 \\
$[$NeIII]~$\lambda 3869$  	& \ldots  & \ldots  & 19.4 $\pm$ 2.8 & \ldots  & \ldots  \\
HeI$+$H8~$\lambda 3889$  	& \ldots  & \ldots  & 27.1 $\pm$ 3.4 & \ldots  & \ldots  \\
H$\epsilon$~$\lambda 3970$  	& \ldots  & \ldots  & 30.0 $\pm$ 3.7 & \ldots  & \ldots  \\
H$\delta$~$\lambda 4101$  	& 117 $\pm$ 41 & \ldots  & 57.2 $\pm$ 6.2 & \ldots  & \ldots  \\
H$\gamma$~$\lambda 4340$  	& 188 $\pm$ 27 & \ldots  & 99.0 $\pm$ 10.2 & \ldots  & 174 $\pm$ 40 \\
H$\beta$~$\lambda 4861$  	& 478 $\pm$ 51 & 73 $\pm$ 12 & 229 $\pm$ 24 & 656 $\pm$ 12 & 376 $\pm$ 45 \\
$[$OIII]~$\lambda 4959$  	& 55 $\pm$ 16 & \ldots  & 116 $\pm$ 14 & \ldots  & 153 $\pm$ 28 \\
$[$OIII]~$\lambda 5007$  	& 223 $\pm$ 30 & \ldots  & 342 $\pm$ 36 & 115 $\pm$ 17 & 455 $\pm$ 52 \\
$[$OI]~$\lambda 6300$  		& \ldots  & \ldots  & 13.1 $\pm$ 4.9 & 158 $\pm$ 18 & \ldots  \\
$[$NII]~$\lambda 6548$  	& 161 $\pm$ 23 & 22.8 $\pm$ 7.1 & 23.6 $\pm$ 4.7 & \ldots  & 85 $\pm$ 13 \\
H$\alpha$~$\lambda 6563$  	& 1730 $\pm$ 170 & 277 $\pm$ 29 & 752 $\pm$ 76 & 1660 $\pm$ 170 & 1510 $\pm$ 150 \\
$[$NII]~$\lambda 6584$  	& 581 $\pm$ 62 & 133 $\pm$ 16 & 98 $\pm$ 11 & 1300 $\pm$ 130 & 299 $\pm$ 33 \\
$[$SII]~$\lambda 6717$  	& 308 $\pm$ 36 & \ldots  & 110 $\pm$ 11 & \ldots  & 131 $\pm$ 16 \\
$[$SII]~$\lambda 6731$  	& 226 $\pm$ 28 & \ldots  & 83.6 $\pm$ 8.9 & \ldots  & 125 $\pm$ 16 

\enddata \tablecomments{Rest emission-line fluxes in units of
$10^{-17}$ \ergscm~before extinction correction. Errors include
statistical measurement uncertainties. Column headings designate the
SN names whose host-galaxy line fluxes are presented; see
Table~\ref{sample_table} for corresponding galaxy names. \\}
\label{lines_SNpos_table}
\end{deluxetable}

\clearpage

\begin{deluxetable}{lcccccccl}
\tablewidth{0pt}
\singlespace
\tabletypesize{\scriptsize}
\tablecaption{Derived Central Properties of Host Galaxies of SN Ic (broad)}
\tablehead{\colhead{SN} & 
\colhead{\emph{$M_B$}} &
\colhead{log(O/H)+12} &
\colhead{log(O/H)+12} &
\colhead{log(O/H)+12} &
\colhead{L(H$\alpha$)\tablenotemark{b}} &
\colhead{SFR\tablenotemark{b}} &
\colhead{$n_e$(SII)} &
\colhead{$E(B-V)$} \\
\colhead{} & 
\colhead{[mag]}&
\colhead{KD02\tablenotemark{a}} & \colhead{M91\tablenotemark{a}} &  \colhead{PP04-O3N2\tablenotemark{a}} &
\colhead{[10$^{40}$ \ergs]} & 
\colhead{[\smy]} & 
\colhead{[\cm]}  & 
\colhead{[mag]}
}
\startdata
SN~1997dq & $-$20.1 &   9.15$^{+ 0.36}_{- 0.19}$\tablenotemark{c} & \nodata  & \nodata  &   $>$0.11 &  $>$ 0.01 &  $<$10 &   \nodata \\
SN~1997ef & $-$20.2 &   8.93$^{+ 0.01}_{- 0.01}$ & \nodata  &   8.89$^{+ 0.03}_{- 0.03}$ &   1.2 &   0.09 &           90 &   0.48 \\
SN~2002ap & $-$20.6 &   \centkd $^{+ 0.04}_{- 0.04}$ &  \centm $^{+ 0.05}_{- 0.05}$ &   \centpp $^{+0.05}_{-0.05}$ & \nodata  & \nodata  & \nodata  & \nodata  \\
SN~2003bg & $-$17.5 &   9.03$^{+ 0.03}_{- 0.03}$\tablenotemark{c} & \nodata  & \nodata  & $> 10^{-4}$ & $> 10^{-5}$  & \nodata  & \nodata  \\
SN~2003jd & $-$20.3 &   8.80$^{+ 0.02}_{- 0.03}$ & \nodata  &   8.54$^{+ 0.03}_{- 0.03}$ &   36.3 &   2.9 &          110 &   0.32 \\
SN~2005da & $-$21.4 &  \nodata & \nodata & \nodata & \nodata  & \nodata  & \nodata  & \nodata  \\
SN~2005fk & $-$21.0 &  \nodata & \nodata & \nodata & \nodata  & \nodata  & \nodata  & \nodata  \\
SN~2005kr\tablenotemark{d} & $-$17.4 &   8.63$^{+ 0.03}_{- 0.03}$ &   8.64$^{+ 0.01}_{- 0.01}$ &   8.24$^{+ 0.01}_{- 0.01}$ &   2.14 &   0.17 &           $<$10 &   0.10 \\
SN~2005ks\tablenotemark{d} & $-$19.2 &   8.87$^{+ 0.02}_{- 0.03}$ &   8.71$^{+ 0.03}_{- 0.04}$ &   8.63$^{+ 0.01}_{- 0.01}$ &  14.0 &   1.11 &           80 &   0.26 \\
SN~2005kz & $-$20.9 &  \nodata & \nodata & \nodata & \nodata  & \nodata  & \nodata  & \nodata  \\
SN~2006nx\tablenotemark{d}  & $-$18.9 &   8.53$^{+ 0.10}_{- 0.14}$ &   8.48$^{+ 0.09}_{- 0.10}$ &   8.24$^{+ 0.04}_{- 0.03}$ &   5.48 &   0.43 & \nodata  &   0.40 \\
SN~2006qk\tablenotemark{d} & $-$17.9 &   8.82$^{+ 0.04}_{- 0.05}$ &   8.61$^{+ 0.07}_{- 0.08}$ &   8.75$^{+ 0.02}_{- 0.03}$ &   7.88 &   0.62 &           $<$10 &   0.52 \\
SN~2007I\tablenotemark{d} & $-$16.9 &   8.71$^{+ 0.03}_{- 0.04}$\tablenotemark{c} & \nodata  &   8.39$^{+ 0.40}_{- 0.17}$ &   0.30 &   0.02 &          290 &   0.34

\enddata
\tablenotetext{a}{Extinction-corrected oxygen abundances derived using the calibrations 
of Kewley \& Dopita (2002, KD02), of McGaugh (1991, M91), and of Pettini
\& Pagel (2004), using the [O III]/[N II] calibration (PP04-O3N2). See \S~\ref{metal_subsec} for details. \\}
\tablenotetext{b}{Extinction-corrected values. Lower limits are indicated if they could not be extinction corrected (when H$\beta$ could not be observed). \\}
\tablenotetext{c}{Oxygen abundance computed using the NII/H$\alpha$ method. \\}
\tablenotetext{d}{Central spectra represent also spectra taken at the SN position since the SN occurred in the galaxy center. See Table~\ref{sample_table}. \\ }

\label{prop_table_nuc}
\end{deluxetable}

\begin{deluxetable}{lcccccccl}
\singlespace
\tabletypesize{\scriptsize}
\tablewidth{0pt}
\tablecaption{Derived Properties of Host Galaxies of SN~Ic (broad) at SN Position if Different from Galaxy Center}
\tablehead{\colhead{SN} & 
\colhead{\emph{$M_B$}} &
\colhead{log(O/H)+12} &
\colhead{log(O/H)+12} &
\colhead{log(O/H)+12} &
\colhead{L(H$\alpha$)\tablenotemark{b}} &
\colhead{SFR\tablenotemark{b}} &
\colhead{$n_e$(SII)} &
\colhead{$E(B-V)$} \\
\colhead{} & 
\colhead{[mag]}&
\colhead{KD02\tablenotemark{a}} & \colhead{M91\tablenotemark{a}} &  \colhead{PP04-O3N2\tablenotemark{a}} &
\colhead{[10$^{40}$ \ergs]} & 
\colhead{[\smy]} & 
\colhead{[\cm]}  & 
\colhead{[mag]}
}
\startdata
SN~1997dq & $-$20.1 &   8.95$^{+ 0.37}_{- 0.21}$\tablenotemark{c} & \nodata  & \nodata  & \nodata  & \nodata  & \nodata  & \nodata  \\
SN~1997ef & $-$20.2 &   8.93$^{+ 0.04}_{- 0.04}$ &   8.84$^{+ 0.02}_{- 0.03}$ &   8.69$^{+ 0.02}_{- 0.03}$ &   0.92 &   0.07 &           70 &   0.24 \\
SN~1998ey & $-$21.8 &   9.08$^{+ 0.04}_{- 0.04}$ & \nodata  & \nodata  &  $>$0.31 & $>$0.02 & \nodata  &   0.12 \\
SN~2002ap & $-$20.6 &   \poskd $^{+ 0.06}_{- 0.06}$\tablenotemark{d} &   \posm $^{+ 0.07}_{- 0.07}$\tablenotemark{d} &   \pospp $^{+ 0.06}_{- 0.06}$\tablenotemark{d} & \nodata  & \nodata  & \nodata  & \nodata  \\
SN~2003bg & $-$17.5 &   8.83$^{+ 0.10}_{- 0.10}$\tablenotemark{c} &  \nodata & \nodata & \nodata  & \nodata  & \nodata  & \nodata  \\
SN~2003jd & $-$20.3 &   8.59$^{+ 0.05}_{- 0.05}$ &   8.62$^{+ 0.02}_{- 0.02}$ &   8.39$^{+ 0.02}_{- 0.02}$ &   0.83 &   0.07 &          120 &   0.14 \\
SN~2005nb & $-$21.3 &   8.65$^{+ 0.08}_{- 0.09}$ &   8.56$^{+ 0.03}_{- 0.03}$ &   8.49$^{+ 0.03}_{- 0.03}$ &  4.33 &   0.34 &          480 &   0.34
\enddata
\tablenotetext{a}{Extinction-corrected oxygen abundances derived using the calibrations 
of Kewley \& Dopita (2002, KD02), of McGaugh (1991, M91), and of Pettini
\& Pagel (2004), using the [O III]/[N II] calibration (PP04-O3N2). See \S~\ref{metal_subsec} for details. \\}
\tablenotetext{b}{Extinction-corrected values. Lower limits are indicated if they could not be extinction corrected (when H$\beta$ could not be observed). \\}
\tablenotetext{c}{Extrapolated using central oxygen abundances from Table~\ref{prop_table_nuc} and assuming standard metallicity gradient. See text for details.\\}
\tablenotetext{d}{Extrapolated using central oxygen abundances from Table~\ref{prop_table_nuc} and measured M~74 metallicity gradient. See text for details.\\}
\label{prop_table_atSNpos}
\end{deluxetable}

\begin{deluxetable}{lccccccl}
\singlespace
\tabletypesize{\scriptsize}
\tablewidth{0pt}
\tablecaption{Derived Properties of Host Galaxies of SN Ic (broad \& GRB) at GRB Position }
\tablehead{\colhead{GRB/SN Name} & 
\colhead{\emph{$M_B$}} &
\colhead{log(O/H)+12} &
\colhead{log(O/H)+12} &
\colhead{log(O/H)+12} &
\colhead{L(H$\alpha$)\tablenotemark{b}} &
\colhead{SFR\tablenotemark{b}} &
\colhead{$n_e$(SII)} \\
\colhead{} & 
\colhead{[mag]}&
\colhead{KD02\tablenotemark{a}} & \colhead{M91\tablenotemark{a}} &  \colhead{$T_e$\tablenotemark{a}} &
\colhead{[10$^{40}$ \ergs]} & 
\colhead{[\smy]} & 
\colhead{[\cm]}  
}
\startdata
\snbw & $-$17.6 &    8.5 &    8.3 &   \ldots &   4.14 &   0.19 & \nodata  \\
\snlw & $-$19.3 &    8.1 &    8.0 &    7.8\tablenotemark{c} & 217.0 &   9.20 & 300 \\
020903 & $-$18.8 &    8.3 &    8.3 &    8.0\tablenotemark{c} &   3.76 &   0.16 & \nodata  \\
\snaj & $-$15.9 &    8.1 &    8.0 &    7.5\tablenotemark{c} &   0.82 &   0.03 & \nodata  \\
\sndh & $-$16.5 &    8.0 &    7.9 &    7.5\tablenotemark{d} &   3.00 &   0.13 & \nodata 
\enddata
\tablenotetext{a}{Extinction-corrected oxygen abundances derived using the calibrations 
of Kewley \& Dopita (2002, KD02), of McGaugh (1991, M91), and in the $T_e$ scale. See \S~\ref{metal_subsec} for details. \\}
\tablenotetext{b}{Extinction-corrected values.}
\tablenotetext{c}{From $T_e$ method and measured [O III]~$\lambda4363$.}
\tablenotetext{c}{From the KD02-$T_e$ conversion using Equation 3 in \citet{kewley07}.}
\label{GRBSN_prop_table}
\end{deluxetable}
\end{document}